\documentclass[prb,aps,twocolumn,groupaddress,nofootinbib,floatfix]{revtex4-2}
\usepackage{latexsym}
\usepackage{hyperref}
\usepackage{times}

\usepackage{amsmath}
\usepackage{multirow} 
\usepackage{dcolumn}
\usepackage{bm}
\usepackage{textcomp}
\usepackage{color}
\usepackage{amssymb}
\usepackage{xcolor}
\usepackage{easyReview}
\usepackage{mathdots}
\usepackage{float}
\usepackage{lineno}
\usepackage{todonotes}
\usepackage{array}
\usepackage{physics}
\usepackage{graphicx}
\usepackage{subfigure}

\usepackage{comment}
\newcommand{\beq}{\begin{eqnarray}}
\newcommand{\eeq}{\end{eqnarray}}

\newcommand{\Cd}{\hat{c}^{\dag}}

\newcommand{\C}{\hat{c}} 
\newcommand{\A}{\hat{a}} 

\begin{document}
\title{Charge Susceptibility and Kubo Response in Hatsugai-Kohmoto-related Models}
\author{ Yuhao Ma$^{1}$, Jinchao Zhao$^{1}$, Edwin W. Huang$^{2,3}$, Dhruv Kush$^{1}$, Barry Bradlyn$^{1}$, Philip W. Phillips$^{1}$}

\affiliation{$^1$Department of Physics and Institute of Condensed Matter Theory, University of Illinois at Urbana-Champaign, Urbana, IL 61801, USA}

\affiliation{$^2$Department of Physics and Astronomy, University of Notre Dame, Notre Dame, IN 46556, United States}

\affiliation{$^3$Stavropoulos Center for Complex Quantum Matter, University of Notre Dame, Notre Dame, IN 46556, United States}

\begin{abstract}

We study in depth the charge susceptibility for the band Hatsugai-Kohmoto (HK) and orbital (OHK) models. As either of these models describes a Mott insulator, the charge susceptibility takes on the form of a modified density response function with lower and upper Hubbard bands, thereby giving rise to a multi-pole structure. The particle-hole continuum consists of hot spots along the $\omega$ vs $q$ axis arising from inter-band transitions.  Such transitions, which are strongly suppressed in non-interacting systems, obtain here because of the non-rigidity of the Hubbard bands.  This modified density response function gives rise to a plasmon dispersion that is inversely dependent on the momentum, resulting in an additional contribution to the conventional f-sum rule.  This extra contribution originates from a long-range diamagnetic contribution to the current. This results in a non-commutativity of the long-wavelength ($q\rightarrow 0$) and thermodynamic ($L\rightarrow\infty$) limits.  When the correct limits are taken, we find that the Kubo response computed with either open or periodic boundary conditions yields identical results that are consistent with the continuity equation contrary to recent claims.   We also show that the long wavelength pathology of the current noted previously also plagues the Anderson impurity model interpretation of dynamical mean-field theory (DMFT). 
\end{abstract}

\maketitle

\section{Introduction}

Spin-spin or density-density response functions encrypt all collective phenomena in strongly correlated matter.  For the charge collective degrees of freedom, the density-density response function, $\chi(\mathbf q,\omega)$, suffices.  While the density-density response encodes the energy range for particle-hole excitations, namely the particle-hole continuum, it also describes more collective excitations involving all the electrons.  The simplest of such excitations are plasma oscillations.  The plasma frequency emerges as the zero of the dielectric response,
\begin{equation}
\epsilon(\mathbf q,\omega) = [1+V_{\mathbf q}\chi(\mathbf q ,\omega)]^{-1},    
\end{equation}
where, $V_{\mathbf q}$ is the Fourier transform of the long-range Coulomb interaction. For systems governed by microscopic Hamiltonians with nonrelativistic kinetic energy and only the Coulomb interaction, the large frequency asymptotic form of $\chi(\mathbf{q},\omega)$ is fixed by the standard f-sum rule to be~\cite{kadanoff1963hydrodynamic} 
\beq\label{eq:chi_asymptotic_ordinary}
\lim_{\omega\rightarrow\infty}\chi(\mathbf q,\omega)\sim q^2 n_e/(m\omega^2).
\eeq
Coupled with the fact that in $d=3$, $V_{\mathbf q}=4\pi e^2/q^2$, the dielectric function reduces exactly to 
\begin{align}
\epsilon({\mathbf q} ,\omega\rightarrow\infty)&\sim\left[1+\left(\frac{\omega_p}{\omega}\right)^2\right]^{-1} \nonumber \\
&\sim 1-\left(\frac{\omega_p}{\omega}\right)^2,
\label{diel}
\end{align}
where $\omega_p = 4\pi e^2 n_e/m$ is the 3d plasma frequency.  When the Coulomb interaction is treated within the random phase approximation (RPA), we can trust this approximation down to $\omega \approx \omega_p$. 
A zero of the dielectric function precisely at $\omega_p$ leads to the plasma oscillations. It is the cancellation of the $q^2$ factors in the asymptotic expansion for the density-density response function and the Coulomb interaction that leads to the pole structure of the dielectric function at $\omega=\omega_p$.  Although in 2d, $V_{\mathbf q}\propto 1/q$ and the asymptotic expansion still scales as $\chi\propto q^2$, just as in 3d.

Furthermore, the f-sum rule that leads to Eq.~\eqref{eq:chi_asymptotic_ordinary} can be generalized beyond simple nonrelativistic Hamiltonians.  In terms of the charge density operator
\beq
\rho_{\mathbf q}=\sum_{\mathbf p,\sigma} c^\dagger_{\mathbf{p},\sigma}c_{\mathbf {{p+q}},\sigma},
\eeq
where $c_{\mathbf p}$ represents the annihilation operator for a fermion with momentum $\mathbf p$,  one can show~\cite{husain2023pines,bradlyn2024spectral} that the f-sum rule
\beq
\int^\infty_{-\infty} \dd{\omega} \omega \Im \chi(\mathbf q ,\omega)=\frac{\pi}{N}\langle[[H,\rho_{\mathbf q}],\rho_{\mathbf q}^\dagger] \rangle
\label{eq:double_commutator}
\eeq
can be recast a double commutator directly from the continuity equation. The double commutator is proportional to the average of the diamagnetic current~\cite{mckay2024charge}.  For a nonrelativistic system with a parabolic band dispersion and momentum-independent interactions, the RHS of Eq. (\ref{eq:double_commutator}) simplifies to $-\pi q^2 n_e/m$. Moreover, this result can be generalized to any band structure.  The long wavelength response of a free electron gas in the presence of Coulomb interactions is tethered to charge conservation via the density-density response function.

Precisely how all of the basic features of the dielectric response change for a doped Mott insulator such as the cuprates is unknown.  The most prominent treatment\cite{gull} of the charge susceptibility in a doped Mott insulator is a numerical study using the dynamical cluster approximation (DCA).  However, this work was performed in imaginary frequency and analytical continuation is necessary to compare with the results we report here.  Where possible, we make a comparison with this work.
Two key questions arise: 1) how the particle-hole continuum is transformed in the presence of the non-rigid lower and upper Hubbard bands, 2) whether the cancellation between the factors of momentum that leads to Eq. (\ref{diel}) still holds if strong correlations between the electrons were included, and whether the asymptotic expansion Eq.~\eqref{diel} remains valid for $\omega$ near $\omega_p$.
If not, is this reflected in the f-sum rule (perhaps restricted to frequencies below the Hubbard gap)?   Only recently\cite{abbamonte1,abbamonte2,abbamonte3} with the advent of momentum-resolved electron-energy loss spectroscopy (MEELS) has $\chi(\mathbf q ,\omega)$ become experimentally accessible.  Experiments on Bi$_{2.1}$Sr$_{1.9}$CaCu$_2$O$_{8+x}$ (BSCCO) reveal that the imaginary part of the density-density response function, $\chi^{''}(\mathbf q ,\omega)$, in the energy range $0.1-2\mathrm{eV}$, consists of a flat temperature and momentum-independent continuum that persists to the eV scales.  At the lowest energy scales, $\chi^{''}\propto q^2$, whereas at high energy it decays as $q^2/\omega^2$.  Also of note is the appearance of a plasmon mode\cite{basov11} that appears to be independent of density. The flat response of $\chi$ at intermediate energy scales and the density independence of the plasma frequency represent dramatic departures from the response in a non-interacting electron gas.

As the cuprates are doped Mott insulators, it is natural to resort to Mottness to resolve these departures from the standard theory of metals.  As the Hubbard model is not solvable in $d=2$, obtaining a rigorous answer is problematic.  Rather than engage in uncontrolled perturbative expansions, we pursue an exactly solvable model for a doped Mott insulator.  Over the last few years\cite{tenkila2024dynamical, dmitry2024, guerci2024electrical, HKfixedp, mai2024, hatsugai1992exactly, skolimowski2024real, phillips2020exact, zhao2023friedel, wang2024simple, mai20231, huang2022discrete, mai2023topological, PhysRevB.97.195102, yeo2019local, setty2024symmetry, wang2023nonlocal, setty2020pairing, zhu2021topological, zhong2024notes, PhysRevB.108.085106, jablonowski2023topological, PhysRevB.109.115108, PhysRevB.103.024529, yang2023bose, jablonowski2023topological, wang2023non, wysokinski2023quantum}, the Hatsugai-Kohmoto model (HK) has been pursued as an exactly solvable platform for Mott physics.  However, in none of these works has the density-density response been investigated.  Three key features emerge in our analysis of the density-density response function of the HK model precisely because the Hubbard bands are not rigid:  1)  the plasma frequency diverges as $1/q$, 2) the particle-hole continuum has a hot spot (divergence) precisely at the frequency corresponding to the energy difference between the upper and lower Hubbard bands, and 3) the current operator contains a non-locality resulting in an extra contribution to the f-sum rule anticipated from the unconventional form of the plasmon dispersion.   

Our manuscript is organized as follows.  In sec.\ref{sec:HK_model}, we briefly overview the HK model. In sec.(\ref{sec:chi}), we derive and evaluate the charge susceptibility for the band HK model.  This computation is done exactly.  However, to couple to an electromagnetic (Maxwell) field to compute the dielectric function, we must also take into account the electric field generated by the electrons due to the Coulomb interaction. We include  the Coulomb interaction self-consistently via the RPA to do so.  Our treatment here then is precisely what is done in Fermi liquid theory (FLT).  In both HK\cite{HKfixedp} and FLT, fixed points exist around which short-range interactions are irrelevant.  No such fixed point has been shown to exist for either FLT or HK in the presence of the long-range Coulomb interaction.  Nonetheless,  motivated by the FL treatment, we treat the Coulomb interaction perturbatively within the RPA. In linear response theory, the system responds not to the applied probe potential $V_{\text{ext}}(q,\omega)$---which determines the density response response of the system---but to the total potential including the internal potential due to the charge of the electrons. The total potential can be written as $V_{\text{tot}}=V_{\text{ext}}(q,\omega)/\epsilon(q,\omega)$, where $\epsilon(q,\omega)$ is the dielectric function
. Thus, near zeroes of the dielectric function the total potential $V_{\text{tot}}(q,\omega)$ diverges, resulting in an enhanced or ‘resonant’ response. The set of $(q, \omega)$  for which $\epsilon(q,\omega)$ is zero then defines the plasmon dispersion.  The only approximation in this work then amounts using RPA to include the coupling to the Maxwell field.  Then we compute the plasma frequency, the f-sum rule and the particle-hole continuum.  Finally, in sec.~\ref{sec:current} we show how one can still obtain a valid definition of the current operator by taking into consideration the non-commutativity of $q\rightarrow 0$ and $L\rightarrow \infty$, thereby rectifying a recent paper\cite{guerci2024electrical} which asserted that such a procedure does not exist although no mention of this pathology was made.  The essence of this non-commutativity can be seen by examining the difference

\beq
\ev{n_{\mathbf k \downarrow} n_{\mathbf k \uparrow}} -  \ev{n_{\mathbf{{k+q}}\downarrow} n_{\mathbf k \uparrow}}
    \label{eqcommutator}
 \eeq

that appears when evaluating the double commutator in Eq. (\ref{eq:double_commutator}). This expression vanishes if we set $q=0$, the required result.  However, for $q\ne 0$, the average values can be factorized because the HK model does not mix the momenta.  As a result, Eq. (\ref{eqcommutator}) will tend to a constant.  Because the latter limit corresponds to the thermodynamic limit, we see that order of limits matters and the order of $q\rightarrow 0$ and $L\rightarrow\infty$ cannot be  exchanged.  This is a subtlety that arises from the long range nature of the interactions in the HK model.   Finally, we discuss the intimate relationship between DMFT and band HK.

\section{Hatsugai-Kohmoto models} \label{sec:HK_model}

We start with the band version of the HK model (bHK)\cite{hatsugai1992exactly} in which there is only one orbital per unit cell, compared to its n-orbital orbital generalization\cite{dmitry2024,mai2024}. The Hamiltonian of the band HK model is
\begin{equation}\label{eq:bHK_hamiltonian}
    H_{\text{bHK}} = \sum_{\mathbf k  \sigma} \xi_{\mathbf k} n_{\mathbf k \sigma} + \sum_{\mathbf k} U n_{\mathbf k \uparrow} n_{\mathbf k \downarrow},
\end{equation}
where $\xi_{\mathbf k}=\epsilon_{\mathbf k}-\mu$ is the kinetic term and $U$ is the momentum-independent density-density repulsion term due to the fixed point structure of this model\cite{HKfixedp}
.
The (Matsubara) Green function of the band HK model has a two-pole structure

\beq
\label{eq:HK_Green}
    G_{\mathbf k \sigma}(i\omega_n) =  \frac{1-\left< n_{\mathbf k \Bar{\sigma}}\right>}{i\omega_{n} - \xi_{\mathbf k }^L} + \frac{\left< n_{\mathbf k \Bar{\sigma}}\right>}{i\omega_{n} - \xi_{\mathbf k } ^ U} ,
\eeq
where $\omega_n=(2n+1)\pi T$ is the fermionic Matsubara frequencies, and the $L/U$ superscript denotes the lower/upper Hubbard band, $\xi^{L}_{\mathbf k } = \xi_{\mathbf k}$, $\xi^{U}_{\mathbf k } = \xi_{\mathbf k} +U $. 

The orbital version\cite{dmitry2024,mai2024} of
the HK model (OHK) includes dynamics through the non-commutativity of the kinetic and interaction terms. The Hamiltonian with $n$ orbitals per unit cell is given by
\begin{equation}
    \begin{aligned}
 H_{\text{OHK}}=& \sum_{\mathbf k,\alpha,\alpha',\sigma}g_{\alpha,\alpha'}({\mathbf k})c_{{\mathbf k}\alpha\sigma}^\dagger c_{{\mathbf k}\alpha'\sigma}-\mu\sum_{{\mathbf k},\alpha}n_{{\mathbf k}\alpha,\sigma} \\&+\sum_{{\mathbf k},\alpha,\alpha'} U_{\alpha,\alpha'} n_{{\mathbf k}\alpha\uparrow} n_{{\mathbf k}\alpha'\downarrow},
 \label{ohk}
 \end{aligned}
\end{equation}
where we use $\alpha,\alpha'=1,\cdots,n$ to label the orbital indices. It has been shown\cite{mai2024} that this model converges rapidly to the Hubbard model as $1/n^{2d}$. Consequently, in $d=\infty$ all the fluctuations vanish for any $n>1$, thereby making the $n=1$ or band HK the exact representation of the Hubbard model in $d=\infty$.  We will revisit this fundamental correspondence when we formulate the current operator, one of the goals of this paper. 

\section{density-density response}\label{sec:chi}

In this section, we explicitly derive the corresponding density response function for the band HK model. Just as in non-interacting systems, the susceptibility is calculated through the density-density correlation function at finite temperature $T=1/\beta$. 
The correlation function can be decomposed into a convolution of two Green functions,
\begin{equation}
\label{chig}
\begin{split}
    \chi^0_{\text{HK}}&(\mathbf q\ne0 ,\tau)  = -\frac{1}{N} \left< T_{\tau}\left[ \rho_{\mathbf q }(\tau) \rho_{\mathbf{-q}}(0) \right]\right>\\
    & = -\frac{1}{N} \left< T_{\tau}\left[ \sum_{\mathbf k} \Cd_{\mathbf k}(\tau) \C_{\mathbf {k+q}}(\tau) \sum_{\mathbf p} \Cd_{\mathbf {p+q}} \C_{\mathbf p} \right]\right>\\
    & = 
    -\frac{1}{N} \sum_{\mathbf {k,p}}\left< T_{\tau}\left[ \Cd_{\mathbf k}(\tau)\C_{\mathbf p} \right]\right> \left< T_{\tau}\left[ \C_{\mathbf {k+q}}(\tau)\Cd_{\mathbf {p+q}} \right]\right>,
\end{split}
\end{equation}
where momentum factorization of HK model (valid for nonzero $\mathbf{q}$) was applied in going from the second to the third line. The other contraction vanishes for any $\mathbf{q}\neq 0$. The final term is proportional to $\delta_{\mathbf k,\mathbf p}$ as the HK model lacks momentum mixing. Upon applying the cyclic property of the trace and anti-periodicity, we arrive at the final result,
\beq\label{eq:chi_convolution}
\chi^0_{\text{HK}}(\mathbf q ,\tau) = \frac{1}{N} \sum_{\mathbf k} G_{\mathbf k}(-\tau) G_{\mathbf{{k+q}}}(\tau).
\eeq

Before we directly substitute the HK Green function into Eq. (\ref{chig}), we can gain more physical insight into the final expression by introducing a few notations.
By first introducing the partition function,
\beq
Z_{\mathbf k} = \Tr e^{-\beta H} = 1 +2 e^{-\beta \xi_{\mathbf k}} + e^{-\beta (2 \xi_{\mathbf k} + U)},
\eeq
we define the statistical weights
 in the lower Hubbard bands and upper Hubbard bands
\begin{align}
w^{L}_{\mathbf k } &= \frac{1-\langle n_{\mathbf k \Bar{\sigma}} \rangle}{Z_{\mathbf k}} = \frac{1+ e^{-\beta\xi_{\mathbf k }}}{Z_{\mathbf k}},\\
 w^{U}_{\mathbf k }  &= \frac{\langle n_{\mathbf k \Bar{\sigma}} \rangle}{Z_{\mathbf k}} = \frac{ e^{-\beta\xi_{\mathbf k }} + e^{-\beta(2\xi_{\mathbf k }+U)} }{Z_{\mathbf k}}.
\end{align}
 We also make use of the Fermi functions
\begin{equation}
    \begin{split}
        & f(\xi_{\mathbf k}) = \frac{1}{1+e^{\beta \xi_{\mathbf k }}},
    \end{split}
\end{equation}
from which follow the identities
\begin{equation}
\begin{split}
        & f(\xi^{L}_{\mathbf k}) w^{L}_{\mathbf k } = \frac{e^{-\beta \xi_{\mathbf k }}}{Z_{\mathbf k}},\ f(\xi^{U}_{\mathbf k}) w^{U}_{\mathbf k } = \frac{e^{-\beta(2\xi_{\mathbf k }+U)}}{Z_{\mathbf k}}\text{.}
\end{split}
\end{equation}

\begin{figure}[t!]
\centering

  \includegraphics[width= 1\linewidth]{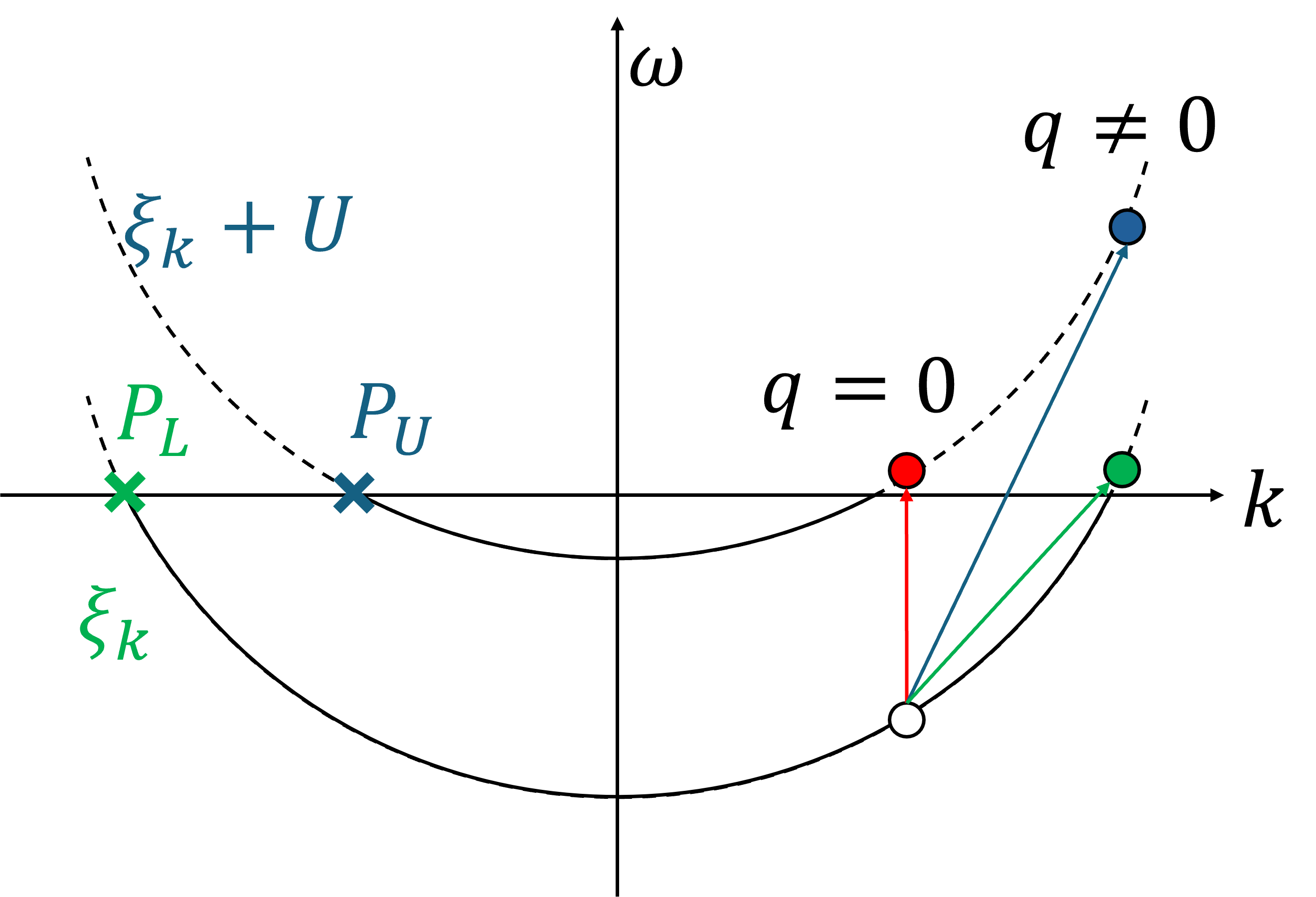}
  \caption{\raggedright  Schematic diagram of filling surface momentum for Hubbard bands $P_L$ and $P_U$. $P_L/P_U$ are the intersections of the chemical potential with the lower/upper Hubble band. The colored arrows represent processes in the particle-hole continuum. Green and blue arrows indicate finite momentum ($q\ne0$) transitions, while the red arrow corresponds to an zero momentum transition which is forbidden in rigid bands. The processes drawn here generalize to the insulating case in which neither the upper nor the lower bands cross the chemical potential.  }

\label{fig:PHC_process}
\end{figure}

These definitions allow us to write the resultant density response susceptibility in Eq.~\eqref{eq:chi_convolution} in the simple form after Fourier transforming and analytically continuing to real frequency:
\beq
\label{eq:HK_lindhard}
     \chi^0_{\text{HK}}(\mathbf q ,\omega) = \frac{1}{N} \sum_{\mathbf k \sigma} \sum_{i,j}^{L,U} w^{i}_{\mathbf {k +q}}  w^{j}_{\mathbf k } \frac{ f(\xi^{j}_{\mathbf k }) - f(\xi^{i}_{\mathbf {k +q}}) }{ \omega + i 0^{+} - \xi^{i}_{\mathbf {k +q}} +  \xi^{j}_{\mathbf k }}.
\eeq
This is the first key result. Eq.~\eqref{eq:HK_lindhard} takes a similar form to the multi-band non-interacting Lindhard function. In the HK system, the lower and upper Hubbard bands comprise the bifurcation of spectral weight (depicted in Fig. (\ref{fig:PHC_process})). 
It has been shown previously\cite{HKfixedp} that the two-pole structure is stable to any local repulsive interactions including Hubbard.  Consequently, the conclusions we reach here regarding the role of the Hubbard bands in the density susceptibility should be quite general.   We can separate the four terms appearing in the sum over $i,j$ in Eq.~\eqref{eq:HK_lindhard} into ``direct'' terms with $(i,j)=(L,L)$ or $(U,U)$, and ``cross'' terms with $(i,j)=(L,U)$ or $(U,L)$. Note that the direct terms involve only a single Hubbard band, while the cross terms involve the mixing of Hubbard bands.

We will use $\mathbf P_i$ to denote the filling surface momentum for Hubbard bands\cite{HKfixedp}, with a Hubbard-band subscript $i=L$ or $U$.   In particular, for an intermediate value of $U<W$, both the upper and lower Hubbard bands cross the chemical potential and we have $\mathbf p_L$ and $\mathbf p_U$ as solutions to
\beq
\xi_{\mathbf p_L} = 0\text{,}\quad \xi_{\mathbf p_U} + U = 0.
\eeq

In the following, we consider  Eq.(\ref{eq:HK_lindhard}) under these approximations:

\begin{enumerate}

 \item Long wavelength and high frequency limit such that $\frac{\mathbf q \cdot\mathbf  P_i }{m(\omega + U)} << 1$. $\omega$ has the unit of energy, while both $\mathbf P_i$ and $\mathbf q$ have the unit of momentum.
 
\item Low/zero temperature limit such that $\partial f / \partial \xi_{\mathbf k} \approx - \delta(\xi_{\mathbf k} )$ where $f$ is the Fermi-Dirac distribution.  In this limit, both spectral weights $w^{L/U}_{\mathbf k }$ are step functions. For $\beta\rightarrow\infty$, we have:
\begin{equation}
\begin{split}
    w^{L}_{\mathbf k } & =    \begin{cases}
    0 &\quad \xi_{\mathbf k } < -U\\
    \frac{1}{2} &\quad -U < \xi_{\mathbf k } < 0 \\
    1 &\quad  \xi_{\mathbf k } > 0 \\
    \end{cases}, \\
    w^{U}_{\mathbf k } & =    
    \begin{cases}
    1 &\quad \xi_{\mathbf k } < -U\\
    \frac{1}{2} &\quad -U < \xi_{\mathbf k } < 0 \\
    0 &\quad  \xi_{\mathbf k } > 0. \\
    \end{cases}
\end{split}
\end{equation}

 \item A parabolic non-interacting dispersion, i.e., $\xi_{\mathbf k} = |\mathbf k|^2 / 2m -\mu$. For clarity, the results are robust even if a tight-binding dispersion is used.

\end{enumerate}

The two direct terms share the same form as the free-electron Lindhard function and thus can be directly evaluated using standard methods\cite{bruus2004many}. The difference of two Fermi functions in the numerator is approximated as a derivative, and then consequently as a delta function at the Fermi level at low temperature. Eventually, for the direct terms we have
\begin{equation}\label{eq:chi_direct}
\chi_{\text{dir}} = \chi_{LL} + \chi_{UU}  = \frac{3 (P_{L}^3 + P_{U}^3) }{32 \pi^2 m } \frac{q^2}{\omega^2}\ +\  \order{q^4}.
\end{equation}
For the cross terms, however, the numerator $f^{L}(\xi_{\mathbf {k +q}}) - f^{U}(\xi_{\mathbf k })$ indicates that it will no longer be just a delta function at the Fermi level. Instead, we shall integrate over a shell with finite thickness proportional to the interaction strength $U$, or the volume of single occupancy. The cross terms,
\begin{equation}\label{eq:chi_cross}
    \chi_{\text{cro}} = \chi_{UL} + \chi_{LU} = \frac{(P_L^3 - P_U^3)U}{12\pi^2(\omega^2 - U^2)}\ +  \order{q^2} ,
\end{equation}
also have a succinct form.  The full susceptibility,
\beq
\chi^0_{\text{HK}} = \chi_{\text{dir}} + \chi_{\text{cro}} 
\eeq
is a sum of the direct and cross terms.  

For Fermi liquids in which $P_L = P_U=P_F$, there is no zeroth-order term, and thus the lowest term would be of $O(q^2)$. Consequently, Eq. (\ref{eq:chi_cross}) comes solely from the mixing of the lower and upper Hubbard bands and is unique to a Mott system. It is worth noting that in the HK model, $\chi$ scales as $q^0$ in both 2D and 3D.  For Fermi liquids,  $\chi$ scales as $q^2$ in both 2D and 3D.  Eq. (\ref{eq:chi_cross}) tells us that the leading cross-term is independent of momentum and is pinned to the frequency $\omega=U$.

Note however, that exactly at $\mathbf{q}=0$, the Lindhard function must vanish for nonzero $\omega$. This obtains because Kubo ensures that $\chi^0_\mathrm{HK}(\mathbf{q},\omega)$ is the Fourier transform of a retarded density-density correlation function, which vanishes identically when $\mathbf{q}=0$, a consequence of particle-number conservation. This implies that Eq.~\eqref{eq:HK_lindhard} should be supplemented with a term proportional to $\delta_{\mathbf{q}0}$ to count for this discontinuity. Since we will be working at small but nonzero $\mathbf{q}$ throughout, this will not affect our main results. However, it does imply that the HK density response function has a removable discontinuity as $\mathbf{q}\rightarrow 0$. As we will see in Sec.~\ref{sec:current}, this discontinuity can be traced to the long-rangedness of the HK interaction in position space. We expect that for a short-range interaction, the discontinuity will be smoothed out, so that $\chi^0(\mathbf{q},\omega)$ still increases rapidly as a function of $\mathbf{q}$ for fixed $\omega$, distinct from its behavior in a Fermi liquid. In that sense, we can view $\chi_{\mathrm{HK}}^0(\mathbf{q},\omega)$ as an exaggerated approximation to the density response of a short-ranged interacting Mott system at finite $\mathbf{q}$.

\section{Plasmon Dispersion}

\begin{figure}[t!]
\centering

  \includegraphics[width= 1\linewidth]{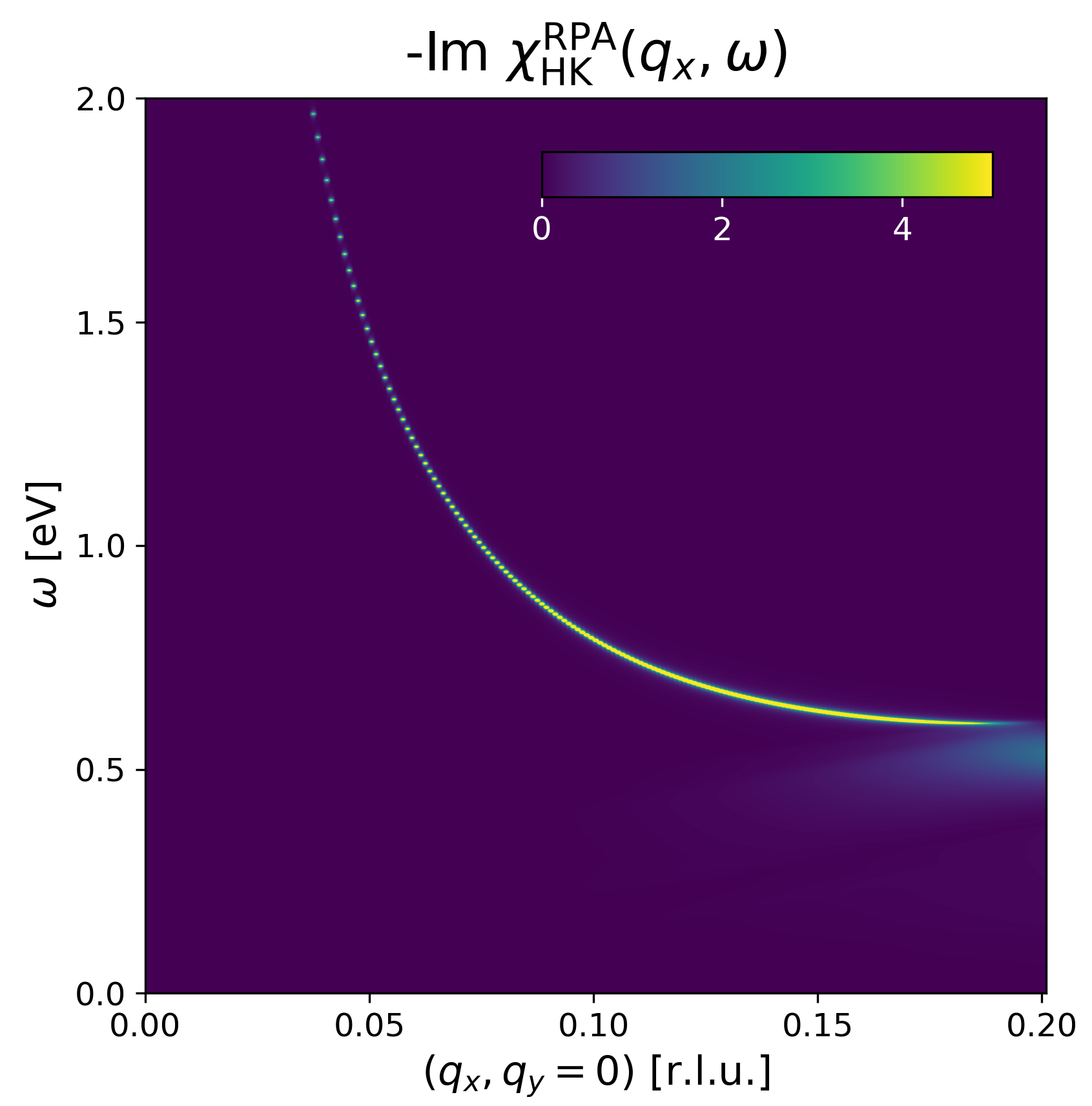}
  \caption{\raggedright Plasmon dispersion in the $(\mathbf{q},\omega)$ plane, as extracted from the pole of $\mathrm{Im} \chi^{\mathrm{RPA}}_{\mathrm{HK}}$. We observe a $1/q$ momentum dependence, suggesting a modification of the f-sum rule from the conventional form in the long wavelength limit. Here, r.l.u. stands for reciprocal lattice units. We work with a square-lattice tight-binding dispersion for our model, $\xi(\mathbf{k}) = -2t\left(\cos(k_x) + \cos(k_y)\right) - 4t_p\cos(k_x)\cos(k_y)$, where $t=0.1$eV, $t_p = 0.03$eV, and $U = 0.14$eV, where $U$ is the interaction strength as defined in Eq. (\ref{eq:bHK_hamiltonian}).
  }

\label{fig:1/q_plasmon}
\end{figure}

Equipped with an analytic expression for the charge susceptibility, we now extract the momentum dependence of the plasma frequency, henceforth referred to as the plasmon dispersion, in the band HK model.
We would now like to compute the plasmon dispersion for the HK model. Thus far we have computed the density response exactly in the HK model which differs substantially from the free electrons as in the Lindhard response.  However, to be able to describe plasmons, we note that  that the HK interaction does not account for the fact that electrons generate an electric field via Maxwell's equations. We can view the HK interaction as an approximation to an exchange interaction mediated by degrees of freedom that have been integrated out to obtain the effective Hamiltonian \eqref{eq:bHK_hamiltonian}. To account for the electric field generated by the charge of the electron, we can consider the Coulomb interaction within RPA using the density-density response function computed in \eqref{eq:HK_lindhard},\eqref{eq:chi_direct}, and \eqref{eq:chi_cross}.   We find that the RPA dielectric function
\beq\label{eq:HK_RPA}
\epsilon^{\text{RPA}}(\mathbf q ,\omega)=1-V_{\mathbf q}\chi^0_{\text{HK}}(\mathbf q ,\omega),
\eeq
and the corresponding RPA density-density response function
\beq
\chi^{\mathrm{RPA}}_{\mathrm{HK}}(\mathbf{q},\omega) = \frac{\chi^0_\mathrm{HK}(\mathbf{q},\omega)}{1-V_\mathbf{q}\chi^0_\mathrm{HK}(\mathbf{q},\omega)}.
\eeq
This does not mean that we have treated the full problem in RPA.  To the contrary, all the Mott physics is encoded into the exact response function, \eqref{eq:HK_lindhard}).  Only the response to the electric field is treated by the RPA but with the HK-Mott physics encoded exactly in $\chi_{\rm HK}^0$.
It is worth noting that here we have chosen a convention such that the imaginary part of susceptibility $\chi^0_{\text{HK}}(\mathbf q ,\omega)$ is negative for both $q>0$ and $\omega>0$, which corresponds to choosing a coupling to the external probe with a positive sign as
\begin{equation}
    \delta H = +\sum_{\mathbf{q}} V_{\mathrm{ext}}(\mathbf{q},t)\rho_{-q}.
\end{equation}
Although the choice of this sign is entirely arbitrary, it does influence the sign of the Coulomb interaction, \( V(\mathbf q ) \), under RPA. An inconsistent choice of signs would result in incorrect position of the pole for the plasmons.

Substituting $V(q)= V_0/q^{2}$ 
into  Eq. \eqref{eq:HK_RPA} for the Coulomb interaction and noting that the direct terms of $\chi(q,\omega)$ scale as $\sim q^{2}/\omega^{2}$ and the cross terms scale as $\sim 1/\omega^{2}$ for small $\mathbf{q}$, we obtain a quartic equation, yielding a solution which scales as
\begin{equation}\label{eq:plasmon_dispersion}
    \omega_{\mathbf p} \sim \frac{1}{q}.
\end{equation}
Thus, we obtain that the plasmon dispersion in band HK is divergent in the $q \rightarrow 0$ limit, as plotted in Fig. \ref{fig:1/q_plasmon}.

 The plasmon dispersion here is unaffected by the removable discontinuity in the density response function at $\mathbf{q}=0$. For a short-ranged interacting system, we would expect the density response function to be continuous as $\mathbf{q}\rightarrow 0$, and so would expect the divergence of the plasma frequency to soften.

Since the conventional $f$-sum rule states that the intensity of a conventional plasmon scales as $q^2$ at small momenta \cite{pines1966}, it is natural to inquire whether such a divergent plasma dispersion of Eq. (\ref{eq:plasmon_dispersion}) would imply a modified form of the $f$-sum rule for HK model. This will be the focus of discussion in the next section, where we will demonstrate that the answer is indeed affirmative. In fact, it is this divergence that reflects a subtlety in defining the current operator.

\section{the f-sum rule}\label{sec:sum_rule}

For non-relativistic noninteracting systems, the plasmon provides the dominant contribution to the $f$-sum rule in the $q\to 0$ limit. Thus, one might expect a modification to the $q$ dependence of the sum rule for the HK model
. In this section, we compute the correction.  To address this, we recall that the standard $f$-sum rule is
\beq\label{eq:f_sum}
\int_{-\infty}^{\infty} \dd{\omega} \omega \Im \chi(\mathbf q , \omega) = - \frac{\pi n_e q^2}{m},
\eeq
where $n_e$ is the average electron density. Note that this equation is derived assuming a free-particle dispersion, i.e., $H = \frac{p^2}{2m}$, and momentum-independent interactions. For a general dispersion, the right-hand side would still scale as $q^2$, though it may exhibit some anisotropy. The integral on the left side can be expressed more generally as a double commutator between the Hamiltonian and the density operator as in Eq. (\ref{eq:double_commutator}).

\begin{figure}[t!]
\centering

  \includegraphics[width= 1\linewidth]{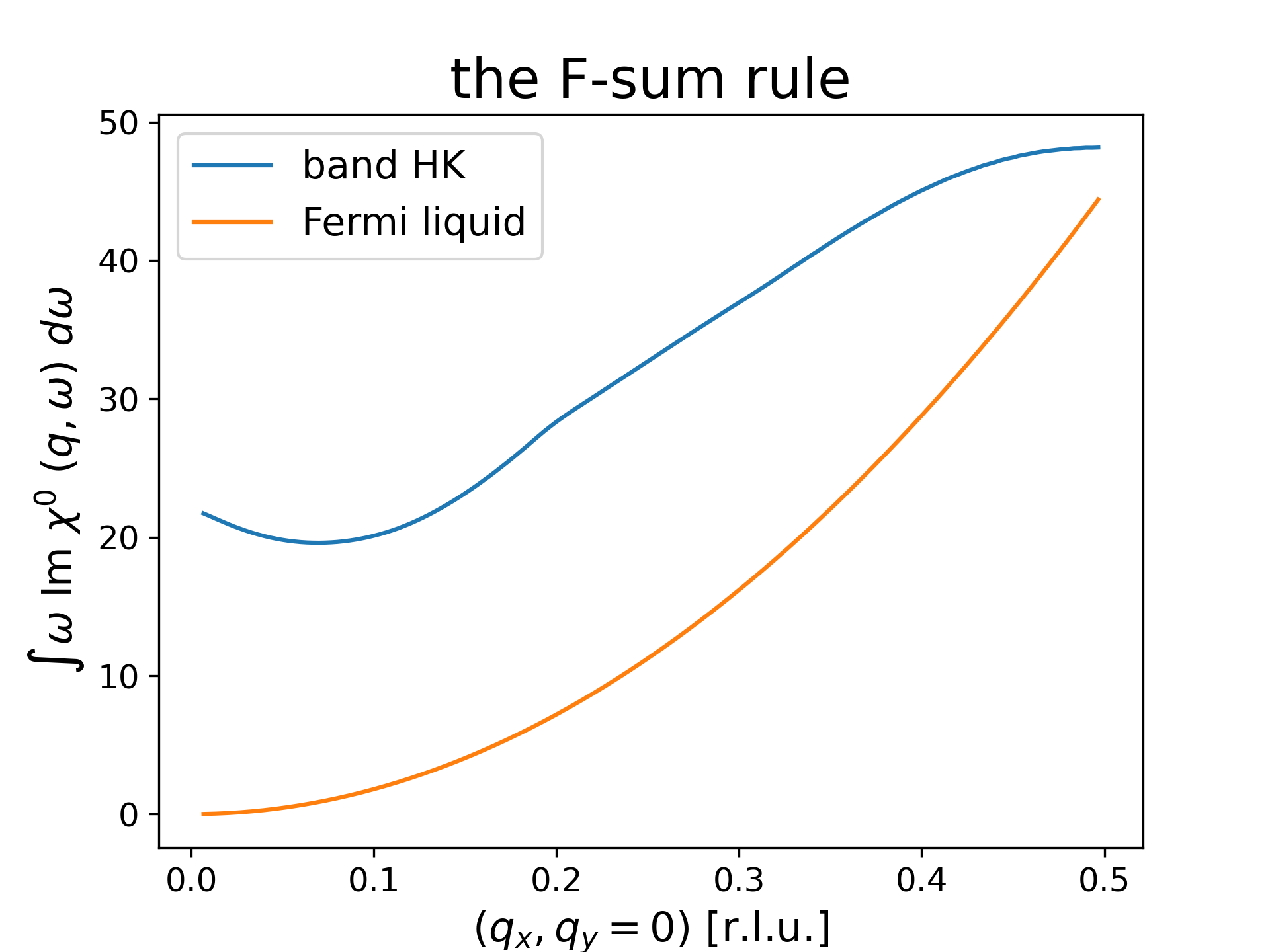}
  \caption{\raggedright The $f$-sum rule for band HK and Fermi liquid systems, plotted as a function of $\mathbf{q}$. The blue line represents the sum rule numerically calculated from the HK model, which clearly does not approach zero as $q \to 0$. For comparison, the orange line represents the sum rule from a generic Fermi liquid, which scales as $q^2$ for small $q$. Note that this plot has the point $q=0$ excluded. The parameters involved in calculating the result for band HK are same as those provided in the caption of Fig.~(\ref{fig:1/q_plasmon}). }

\label{fig:f_sum_comparison}
\end{figure}

One insight we can draw from Eq. (\ref{eq:double_commutator}) is that since the Coulomb interaction commutes with the density operator, Eq. (\ref{eq:f_sum}) holds true for both $\chi^0$ and $\chi^{\text{RPA}}$. Therefore, the intensity of the plasmon, which stems from Coulomb interactions, is also described by the $f$-sum rule and thus scales as $q^2$ at small momenta.
Eq.~(\ref{eq:double_commutator}) generalizes the $f$-sum rule to arbitrary systems. We can use it directly to answer the question of how the HK interaction, $H_{\text{I,bHK}} = \sum_{\mathbf k} U n_{\mathbf k \uparrow} n_{\mathbf k \downarrow}$, may modify the f-sum rule. We find that for the band HK interaction, the double commutator is
\begin{equation}\label{eq:HK_dc}
\begin{split}
     & \ev{\comm{\comm{H_{\text{I,bHK}}}{\rho_{\mathbf q}}}{\rho_{\mathbf q}^\dagger}} = \sum_{\mathbf k}U ( 2  \ev{n_{\mathbf k \downarrow} n_{\mathbf k \uparrow}} \\ 
    & -  \ev{n_{\mathbf{{k+q}}\downarrow} n_{\mathbf k \uparrow}} -  \ev{n_{\mathbf{k-q}\downarrow} n_{\mathbf{k}\uparrow}}) + \qty(\uparrow \leftrightarrow \downarrow),
\end{split}
\end{equation}
which exposes the subtlety in the $q\rightarrow 0$ limit. 
The first term in Eq.~\eqref{eq:HK_dc} represents double occupancy. If we simply set $\mathbf{q}=0$ on the RHS, we obtain that the double commutator vanishes which is the desired result. However, for any $\mathbf{q} \neq 0$, the second terms can be factorized: $\ev{n_{\mathbf{{k+q}} \downarrow} n_{\mathbf{k} \uparrow}} = \ev{n_{\mathbf{{k+q}} \downarrow}} \ev{n_{\mathbf{k} \uparrow}}$ because each $\mathbf{k}$ point in the HK model is decoupled.  If we now use this form to construct the $\mathbf{q}=0$ limit, we would arrive at a non-zero value,
\begin{equation}
\begin{split}
    & \int_{-\infty}^{\infty} \dd{\omega} \omega \Im \chi_{\text{HK}} (\mathbf q, \omega) \\
     = & \frac{4 \pi U}{N} \sum_{\mathbf k}  \qty(\ev{n_{\mathbf k\downarrow} n_{\mathbf k\uparrow}} - \ev{n_{\mathbf k\downarrow}}\ev{n_{\mathbf k\uparrow}}),
\end{split}
\end{equation}
which tends to a finite value.  This discrepancy is a manifestation of the $\delta_{\mathbf{q}0}$ removable discontinuity in the density response function discussed previously. This is also observed in Ref.~\onlinecite{guerci2024electrical} where the subtlety of  $\mathbf{q}=0$ limit is treated incorrectly and yields undesirable results.

The $f$-sum rule is intrinsically linked to the definition of the current operator $\mathbf{j}(q)$ via the continuity equation
\beq
\comm{H}{\rho_\mathbf{q}} = \mathbf q \cdot \mathbf J (q) .
\eeq
 In particular, it can be shown~\cite{mckay2024charge} using the continuity equation that the double commutator must be given by $q^2$ times the diamagnetic current defined via minimal coupling.  Consequently, this non-traditional form of the $f$-sum rule would lead to a subtlety in defining the current operator,  and also to a nonlocal diamagnetic current. We note that nonlocal diamagnetic currents have previously been considered to explain electrodynamics of strongly correlated metals, so this is not necessarily surprising\cite{la2019colloquium}. Regardless of this problem, it is possible to formulate the current operator that is well-defined in the long-wavelength limit, as shall be discussed in sec.(\ref{sec:current}).

\section{Particle—hole Continuum}
\begin{figure}[t!]
\centering
\subfigure[]{\includegraphics[width=0.8\linewidth]{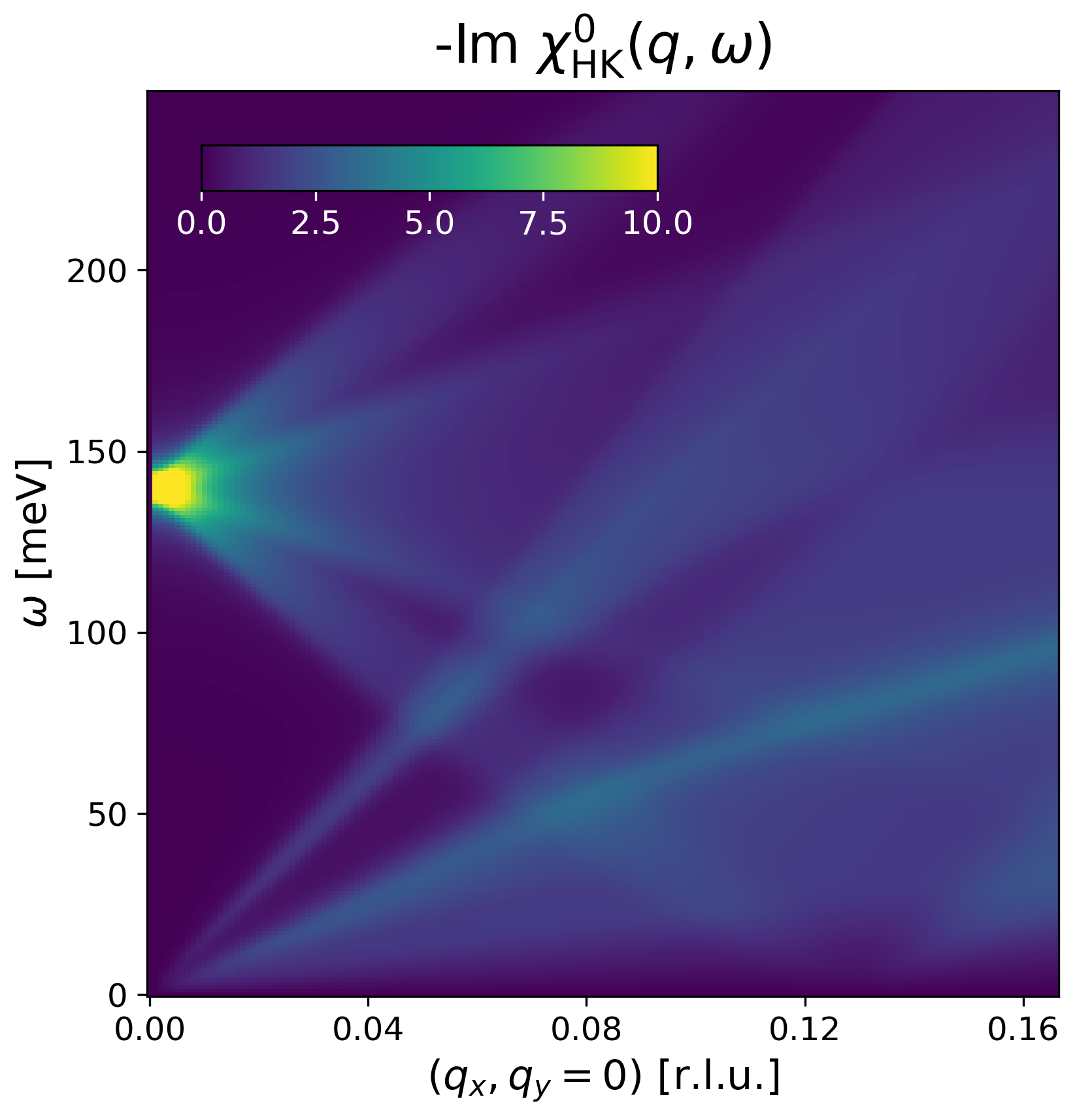}}
\subfigure[]{\includegraphics[width=0.8\linewidth]{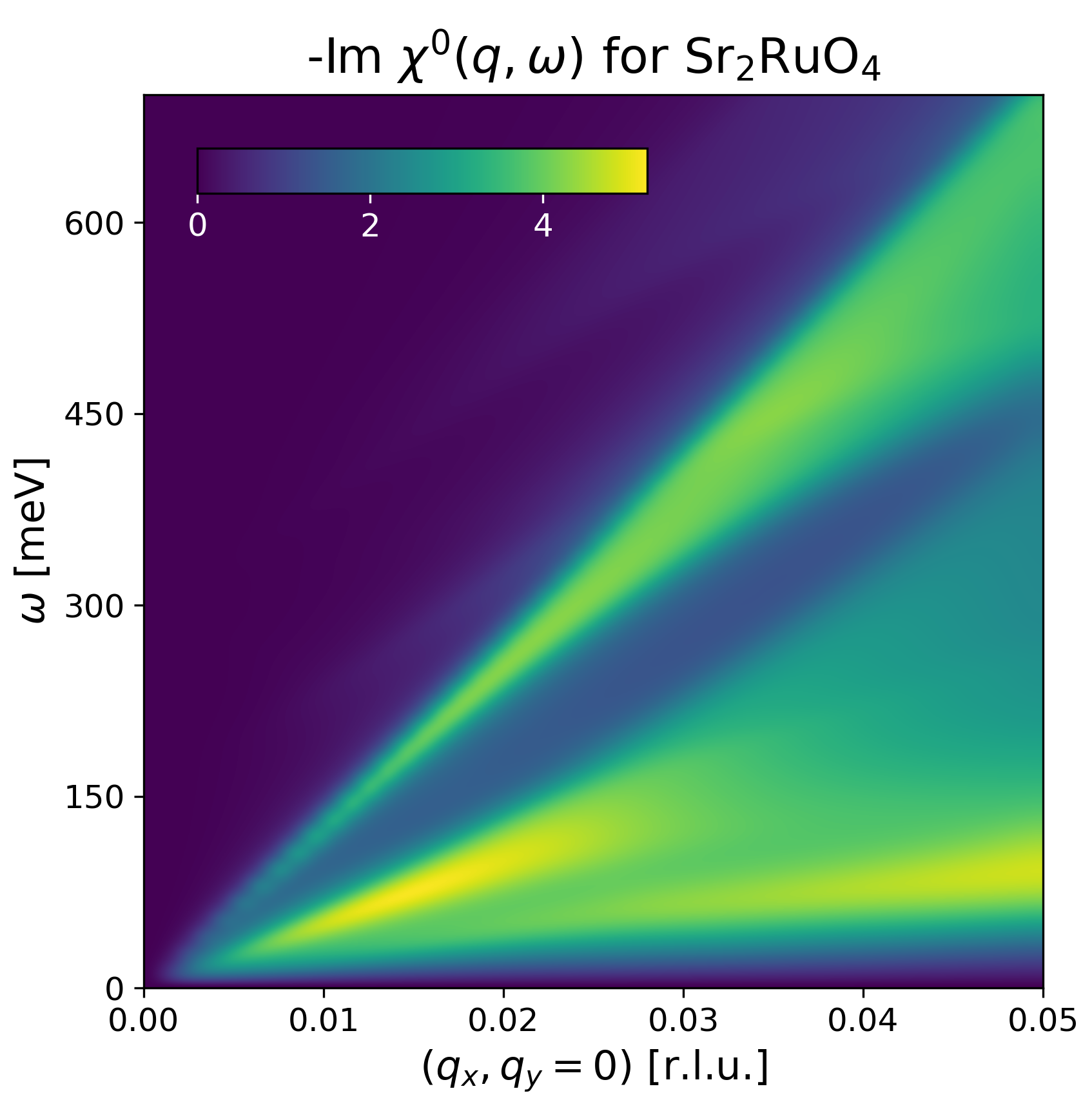}}
  \caption{\raggedright   The imaginary part of the susceptibility for (a) HK model  and (b) Fermi Liquid $\text{Sr}_2\text{Ru}\text{O}_4$. The HK charge susceptibility exhibits a prominent peak along the $\omega$-axis, arising precisely at $\omega=U$, where $U$ is the interaction strength as defined in Eq. (\ref{eq:bHK_hamiltonian}). The high intensity at small momenta indicates that the HK model adheres to a different $f$-sum rule as compared to a Fermi liquid. For the the Fermi Liquid $\text{Sr}_2\text{Ru}\text{O}_4$, the prominent intensities corresponds to the intra-band PHC. The inter-band energy spacing is roughly $\omega = 100$ meV, where the PHC originates on the $\omega$-axis.}

\label{fig:PHC_HK}%
\end{figure}

In this section, we provide an intuitive and straightforward argument to address the underlying physics of all the unconventional phenomena discussed in previous sections, including the divergent plasmon and the unconventional form of the $f$-sum rule. 

It is known that the band HK model exhibits a large ground-state degeneracy \cite{dmitry2024}, which scales as $2^{\Omega_1}$, where $\Omega_1$ represents the number of states in the singly occupied region of the Brillouin zone. This raises the natural question of whether this degeneracy is the cause of the unconventional physics here. We will show that this is not the case. Fig. \ref{fig:PHC_HK}a illustrates the imaginary part of HK susceptibility from Eq. (\ref{eq:HK_lindhard}). In the $\omega-q$ plane, the regions with non-zero intensity represent the particle-hole continuum (PHC), which corresponds to processes whereby an electron absorbs energy $\omega$ and becomes excited to an initially unoccupied state, accompanied by a momentum change of $q$. This results in the creation of a particle-hole pair, as illustrated by the colored arrows in Fig.(\ref{fig:PHC_process}). Unlike the plasmon, this particle-hole excitation is not a collective excitation of electrons.

In a single-band system, or one with a single Fermi surface, the PHC strictly exhibits no intensity near the $\omega$-axis, as processes involving finite energy transfer without a corresponding momentum change do not exist.  That is, all transitions are of the intra-band variety. 
The intra-band PHC originates from the origin and extends along the $q$-axis.
On the other hand, for multi-band systems with more than one Fermi surface, the PHC can extend to nonzero $\omega$ at small $q$, as inter-band transitions can be of finite energy and small momentum. We refer to this as the inter-band PHC.

Despite this, such inter-band processes for non-interacting systems are significantly suppressed in intensity as $q\to 0$. To illustrate this explicitly, let us recall the non-interacting multi-band density response function\cite{lv2013dielectric}:
\begin{align}
       \chi^{0}_{\alpha\beta} & = \frac{1}{N} \sum_{\mathbf k } \sum_{\alpha\beta} \frac{f(\epsilon_{\mathbf k ,\beta }) - f(\epsilon_{\mathbf {k +q},\alpha})}{\hbar \omega + i0^{+} - \epsilon_{\mathbf {k +q},\alpha} + \epsilon_{\mathbf k ,\beta} } \mathcal{F}_{\beta \alpha}(\mathbf k, \mathbf{{k+q}})\text{,}\nonumber\\
    & \mathcal{F}_{\beta \alpha}(\mathbf k, \mathbf{{k+q}}) = \Big| \bra{u_{\beta}(\mathbf k)} \ket{u_{\alpha}(\mathbf{{k+q}})}\Big|^2.
   \end{align}
Here $\alpha$ and $\beta$ are the band indices and $\ket{
u_{\alpha}(\mathbf k)}$ denotes the energy eigenstate with momentum $\mathbf k$ in band $\alpha$. $\mathcal{F}$ thus represents the overlap between two eigenstates.  Consequently, $\mathcal{F}_{\beta \neq \alpha}(\mathbf k, \mathbf q=0) = 0$, $\forall \mathbf k$.  In other words, inter-band transitions are strictly prohibited at $q=0$ because non-interacting bands arise from the diagonalization of the non-interacting Hamiltonian, resulting in static bands. For small $q$, the overlap is no longer zero but remains suppressed, as shown explicitly by
\beq
    \lim_{q \to 0}\mathcal{F}_{\beta \neq \alpha}(\mathbf k, \mathbf{{k+q}}) \approx |\braket{u_{\beta}(\mathbf k)}{\nabla_{\mathbf{k}}u_{\alpha}(\mathbf{k})}|^2 q^2 + \order{q^4}.
\eeq

This suppression ensures that in multi-band non-interacting systems, the intensity of the inter-band PHC scales no faster than $q^2$ at small momenta, in accordance with the $f$-sum rule.  As an example, in Fig. \ref{fig:PHC_HK}b we plot the PHC of $\text{Sr}_2\text{Ru}\text{O}_4$ as obtained from Ref.~\onlinecite{husain2023pines}, which behaves as a Fermi liquid. Its inter-band PHC is faint and barely discernible above the much stronger intra-band PHC. 

However, in the HK density response function, as described in Eq. \ref{eq:HK_lindhard}, the overlap function $\mathcal{F}$ is replaced by the product of two weight functions, $w^{L}_{\mathbf{{k+q}}} w^{U}_{\mathbf{k}}$, which behave as step-like functions at low temperatures. As a result, all electrons within the single-occupancy region are capable of undergoing inter-band transitions of infinitesimal but non-zero momenta $q$ with no suppression, which makes up the peak in Fig. \ref{fig:PHC_HK}, that is, the high intensities of inter-band PHC at $\omega = U$. In the DCA work\cite{gull} , they did not perform the analytical continuation for $q=(0,0)$ and hence the hot-spot we find here could not be studied. We have also checked that if $U$ becomes $\mathbf{k}$-dependent $U_{\mathbf{k}}$ , this hot spot becomes more spread out but in the range of $U_{\mathbf{k}}$, consistent with our argument. In other words, while electrons in non-interacting systems perceive states on every band distinctly, in the HK model they do not, as the upper band simply emerges from double occupancy.  That is, in a Mott system, there is no strict orthogonality between the bands as the bands are non-rigid in contrast to a non-interacting system.

This argument is consistent with the analytical results in Sec.(~\ref{sec:chi}). The direct terms of Eq. (\ref{eq:chi_direct}), which correspond to the intra-band PHC, still preserve the same scaling in $q$ as in non-interacting systems. Only the cross terms of Eq. (\ref{eq:chi_cross}) correspond to the inter-band PHC that has a different scaling in $q$.  Consequently, we identify the non-rigidity of the Hubbard bands---rather than the degeneracy of the HK ground state---as the root cause of the hot spot in the spectral function along the $\omega$ axis.  

Nevertheless, inter-band transitions are also prohibited at strictly $q=0$ in the HK model because it is impossible to excite an electron from a singly-occupied state to a doubly-occupied state at exactly the same $k$-point in momentum space, without either increasing the total particle number by one or involving another electron, thereby turning it into a many-body process. This observation is consistent with the behavior of the double-commutator at $q=0$, where the second and third terms in Eq. (\ref{eq:HK_dc}) are no longer factorizable and thus would cancel with the first term. 
This results in the aforementioned discontinuity in the HK model at $q=0$, which will be explored in detail in the sec.(\ref{sec:current}).

\section{Orbital HK model}

Up to this point, all of our analysis has been based on the band HK model. In this section, we will demonstrate that the main results remain qualitatively unchanged even in the extension to the multi-orbital (OHK).

In the OHK model, the non-commutativity of the interaction and non-interacting Hamiltonians, $\comm{H_{\text{I,OHK}}}{H_0} \neq 0$, has several consequences. First, the occupation number operator $n_{\mathbf k\alpha\sigma}$ is no longer a conserved quantity. Secondly, the time evolution of operators becomes non-trivial. As a result, we must rely on the spectral (Lehmann) representation 
\begin{equation}
    \begin{split}
        \mathcal{G}^{\text{OHK}}_{\alpha\beta}(\mathbf k, \omega+i0^+) & = \frac{1}{Z} \sum_{m,n} \frac{e^{-\beta E_n} + e^{-\beta E_m}}{\omega+i0^+ + E_m - E_n} \\
        & \times \bra{m}  \C_{\mathbf k\alpha\sigma} \ket{n} \bra{n} \Cd_{\mathbf k\beta\sigma} \ket{m},
    \end{split}
\end{equation}
 to calculate the Green function.  Consequently we find for the density response function in orbital HK 
\begin{widetext}
\begin{align}
        & \chi^0_\mathrm{OHK}(\mathbf q,\omega+i0^+ ) \nonumber \\
        & = \frac{1}{N} \sum_{\mathbf k\alpha\beta} \sum_{mn} \sum_{uv} \frac{1}{Z_{\mathbf k} Z_{\mathbf{{k+q}}}} \frac{e^{-\beta(E_m + E_v)} - e^{-\beta(E_n + E_u)}}{\omega+i0^+ + E_n - E_m + E_u - E_v }  \times \bra{m}  \C_{\mathbf k\beta} \ket{n} \bra{n} \Cd_{\mathbf k\alpha} \ket{m} 
        \bra{u}  \C_{\mathbf{{k+q}}\alpha} \ket{v} \bra{v} \Cd_{\mathbf{{k+q}}\beta} \ket{v},
\end{align}
\end{widetext}
where $\sum_{mn} $ and $ \sum_{uv}$ are sums over the complete set of many-body eigenstates. The numerical evaluation of this expression is computationally expensive, as it requires the product of two sets of loops: one loop of length $(2^{2n_\alpha})^4$, where $n_\alpha$ represents the number of orbitals (yielding $16^4 = 65,536$ iterations for a 2-orbital system), and a second loop of length $L_y \times L_x^2 \times n_\alpha^2$ , which accounts for $k$, $q$, and orbital indices. The total number of iterations required is the product of these two quantities.

However, as discussed in Sec. (\ref{sec:sum_rule}), the $f$-sum rule offers a valuable shortcut for gaining insights, particularly through the evaluation of the double commutator in Eq. (\ref{eq:double_commutator}). In particular, when the OHK interaction does not have interactions between different orbitals, such that $H_\text{I,OHK} = \sum_{\mathbf{k} \alpha} U_{ \alpha} n_{\mathbf{k} \alpha \uparrow} n_{\mathbf{k} \alpha \downarrow}$, the double commutator 
\begin{equation}
\begin{split}
    & \ev{\comm{\comm{H_\text{I,OHK}}{\rho_{\mathbf q}}}{\rho_{\mathbf q}^\dagger}} = \sum_{\mathbf k \alpha} \Big( 2 U_{\alpha} \ev{n_{\mathbf k\alpha\downarrow} n_{\mathbf k\alpha\uparrow}} \\
    & - U_{\alpha} \ev{n_{\mathbf{{k+q}},\alpha\downarrow} n_{\mathbf k\alpha\uparrow}} -  U_{\alpha} \ev{n_{\mathbf{k-q},\alpha\downarrow} n_{\mathbf k\alpha\uparrow}} \Big) + \qty(\uparrow \leftrightarrow \downarrow)
\end{split}\label{eq:OHK_DC}
\end{equation}
differs from that of the band HK model in Eq.~(\ref{eq:HK_dc}) only by the inclusion of an additional band index.  As before, we see that as $q\to 0$ the double commutator does not vanish regardless of the number of orbitals. However, in the exact Hubbard limit where the number of orbitals equals the number of sites (which goes to infinity in the thermodynamic limit), Eq.~\eqref{eq:OHK_DC} yields exactly zero on the RHS.  Consequently, we expect that our main results are still valid for OHK and henceforth our analysis in sec.(\ref{sec:current}) will rely on a general form of multi-orbit interaction. Furthermore, this also provides further evidence that the large ground-state degeneracy is irrelevant to the phenomena we are observing, as the OHK model has inherently eliminated this degeneracy \cite{dmitry2024}.

\section{Kubo Response: Current Operator}\label{sec:current}

The divergence of the plasma frequency in the HK model implies that subtleties arise in the definition of the current operator.  In this section we point out 1) how these difficulties can be overcome in HK and 2) that such difficulties arise in the standard formulation of dynamical mean-field theory (DMFT).  

We start by considering a general orbital HK model, which we can write in position space as $H=H_0+H_{\mathrm{I}}$ with
\begin{align}
    H_0 &= \sum_{\substack{\alpha\beta\\\mathbf{RR'}}} c^\dag_{\alpha \mathbf{R}}h^{\alpha\beta}(\mathbf{R-R'})c_{\beta\mathbf{R'}}, \\
    H_\mathrm{I} &= \sum_{\substack{\alpha\beta\\\mathbf{R}_1\mathbf{R}_2\\\mathbf{R}_3\mathbf{R}_4}} U_{\alpha\beta}(\mathbf{R}_1+\mathbf{R}_3-\mathbf{R}_2-\mathbf{R}_4)c^\dag_{\alpha\mathbf{R}_1}c_{\alpha\mathbf{R}_2}c^\dag_{\beta\mathbf{R}_3}c_{\beta\mathbf{R}_4}\label{eq:ohkposition}
\end{align}
where we use $\mathbf{R}_i$ to denote Bravais lattice vectors, and $c_{\alpha\mathbf{R}}$ annihilates an electron in an orbital $\alpha$ at position $\mathbf{R}+\mathbf{r}_\alpha$. 
$U_{\alpha\beta}(\mathbf{R})$ is the orbital HK interaction; canonical anticommutation relations require that $U_{\alpha\beta}(\mathbf{R})$ is symmetric under $\alpha\leftrightarrow\beta$, and vanishes when $\alpha=\beta$. For the case of a momentum-independent orbital HK interaction, $U_{\alpha\beta}(\mathbf{R})\rightarrow U_{\alpha\beta}\delta_{\mathbf{R}\mathbf{0}}$\footnote{Note that for systems with periodic boundary conditions, $\delta_{\mathbf{R}\mathbf{0}}$ should be interpreted as a periodic delta function.}. Our goal will be to compute the interaction contributions to the current operator arising from Eq.~\eqref{eq:ohkposition} and show that the continuity equation is satisfied. While the momentum space continuity equation
\begin{equation}
[H_I,\rho_\mathbf{q}] = \mathbf{q}\cdot\mathbf{j}_{I,\mathbf{q}}\label{eq:interactioncontinuity}
\end{equation}
has been used previously\cite{failure} to determine the interaction contribution $\mathbf{j}_{I,\mathbf{q}}$ to the current operator, Ref.~\onlinecite{guerci2024electrical} has recently cast doubt on this procedure for the long-ranged HK interaction. Here we present an alternative derivation of $\mathbf{j}_{I,\mathbf{q}}$ from minimal coupling which avoids the problem noted in Ref.~\cite{guerci2024electrical}. Our derivation is valid with both open or periodic boundary conditions and shows that the continuity equation Eq.~\eqref{eq:interactioncontinuity} is satisfied for any finite-sized system. This will allow us to identify the origin of the f-sum rule violation in HK with the existence of a long-range diamagnetic contribution to the current.  In essence, we find that in the presence of long-range interactions, $\lim_{q\rightarrow 0}$ and $\lim_{L\rightarrow\infty}$ do not commute. We argue that due to the infinite range of the interaction, a conserved current can be consistently defined for any finite sized system of lenght $L$, and that for every value of $L$ the current has a removable discontinuity as $q\rightarrow 0$. We thus show that conserved charge transport should be obtained by applying the limits in the order 1.) $\lim_{q\rightarrow 0}$ followed by 2.) $\lim_{L\rightarrow\infty}$  rather than the inverse.  It is for this reason that we work with a finite-size system and
take the limit as the system size goes to infinity after evaluating derivatives and commutators.

To show this, we first in subsection~\ref{sec:gauge} derive the gauge-invariant form of the HK interaction in the presence of an external electromagnetic field. By taking variational derivatives of this interaction with respect to the vector potential, we derive the current density operator in position space. Next, in subsection \ref{sec:currentft}, we take the Fourier transform of the current density to derive the interaction contribution to $\mathbf{j}_\mathbf{q}$ in the HK model. We pay particular attention to the subtleties associated with the choice of boundary conditions of the system, and show that for uniform electric fields we recover the observation of Ref.~\onlinecite{failure} that the HK interaction does not contribute to the current operator. Finally, in subsection \ref{sec:thermo} We explore the subtleties associated with taking the thermodynamic limit for the HK model, and show how a careful treatment of the limit allows us to define a conserved, nonsingular current operator as $\mathbf{q}\rightarrow 0$.

\subsection{Gauge-invariant interaction and the current operator}\label{sec:gauge}

To begin, we note that the interaction Eq.~\eqref{eq:ohkposition} is not invariant under local gauge transformations $c_{\alpha\mathbf{R}}\rightarrow e^{i\phi(\mathbf{R-r_\alpha})}c_{\alpha\mathbf{R}}$. This implies that in the presence of a nonvanishing background electromagnetic vector potential $\mathbf{A}(\mathbf{r})$, the interaction Eq.~\eqref{eq:ohkposition} must be modified to preserve gauge invariance. In minimal coupling, the simplest such modification is
\begin{widetext}
\begin{align}
U_{\alpha\beta}(\mathbf{R}_1+\mathbf{R}_3-\mathbf{R}_2-\mathbf{R}_4)&\rightarrow U^A_{\alpha\beta}(\mathbf{R}_1+\mathbf{R}_3-\mathbf{R}_2-\mathbf{R}_4) \nonumber \\
&=U_{\alpha\beta}(\mathbf{R}_1+\mathbf{R}_3-\mathbf{R}_2-\mathbf{R}_4)e^{-i(\int_{\mathbf{R}_1+\mathbf{r}_\alpha}^{\mathbf{R}_2+\mathbf{r}_\alpha}\mathbf{A}(\mathbf{x})\cdot d\mathbf{x} +\int_{\mathbf{R}_3+\mathbf{r}_\beta}^{\mathbf{R}_4+\mathbf{r}_\beta} \mathbf{A}(\mathbf{x})\cdot d\mathbf{x})}\label{eq:mincoupling}
\end{align}
\end{widetext}
This Peierls-like minimal coupling substitution renders the Hamiltonian invariant under the gauge transformation $c_{\alpha\mathbf{R}}\rightarrow e^{i\phi(\mathbf{R-r_\alpha})}c_{\alpha\mathbf{R}}, \mathbf{A}\rightarrow A+\nabla\phi$. It can be viewed as a tight-binding approximation to the minimal coupling of the momentum-dependent HK interaction to the electromagnetic field. Note, however, that there is an ambiguity in Eq.~\eqref{eq:mincoupling}, as we are free to choose any paths we wish to evaluate the line integrals in the Peierls phases. To maintain generality we introduce functions $\mathbf{s}_{ij,\alpha}(\lambda)$ satisfying $\mathbf{s}_{ij,\alpha}(0)=\mathbf{R}_i+\mathbf{r}_\alpha$, and $\mathbf{s}_{ij,\alpha}(1)=\mathbf{R}_j+\mathbf{r}_\alpha$, such that
\begin{equation}
    \int_{\mathbf{R}_1+\mathbf{r}_\alpha}^{\mathbf{R}_2+\mathbf{r}_\alpha}\mathbf{A}(\mathbf{x})\cdot d\mathbf{x} 
    = \int_0^1d\lambda\ \mathbf{s}^\prime_{ij,\alpha}(\lambda)\cdot\mathbf{A}(\mathbf{s}_{ij,\alpha}(\lambda)).
\end{equation}

Now that we know how the interaction minimally couples to the vector potential, we can define the current as a variational derivative of the Hamiltonian with respect to $\mathbf{A}$. Focusing on the interaction contribution, we have
\begin{widetext}
\begin{equation}
\begin{split}
&j_\mathrm{I}^\mu(\mathbf{r}) =  -\left.\frac{\delta H_\mathrm{I}}{\delta A_\mu(\mathbf{r})}\right|_{\mathbf{A}\rightarrow 0} \\
&=i\sum_{\substack{\alpha\beta\\\mathbf{R}_1\mathbf{R}_2\\\mathbf{R}_3\mathbf{R}_4}} U_{\alpha\beta}(\mathbf{R}_1+\mathbf{R}_3-\mathbf{R}_2-\mathbf{R}_4)c^\dag_{\alpha\mathbf{R}_1}c_{\alpha\mathbf{R}_2}c^\dag_{\beta\mathbf{R}_3}c_{\beta\mathbf{R}_4}\times\int_0^1d\lambda s^{\prime,\mu}_{12,\alpha}(\lambda)\delta(\mathbf{r}-\mathbf{s}_{12,\alpha}(\lambda))+s^{\prime,\mu}_{34,\beta}(\lambda)\delta(\mathbf{r}-\mathbf{s}_{34,\beta}(\lambda)) \label{eq:jIposition}
\end{split}
\end{equation}
\end{widetext}

Eq.~\eqref{eq:jIposition} gives the contribution to the local current density coming from the HK interaction and is valid for finite systems with either open or periodic boundary conditions provided the choice of the function $\mathbf{s}_{ij,\alpha}$ is consistent with the boundary conditions. We can now verify that this current is conserved by taking its divergence. For any finite-sized system the sums over position and orbitals in Eq.~\eqref{eq:jIposition} contain a finite number of terms, so we can differentiate term wise. The derivative with respect to $\mathbf{r}$ converts the integrands in Eq.~\eqref{eq:jIposition} into total derivatives with respect to $\lambda$, showing immediately that
\begin{equation}
    \nabla\cdot\mathbf{j}_{I}(\mathbf{r}) = -i[H_I,\rho(\mathbf{r})]\label{eq:continuityposition}
\end{equation}
for \emph{any} choice of $\mathbf{s}_{ij,\alpha}(\lambda)$, where
\begin{equation}
\rho(\mathbf{r}) = \sum_{\alpha\mathbf{R}} \delta(\mathbf{r-R-r_\alpha})c^\dag_{\alpha\mathbf{R}}c_{\alpha\mathbf{R}}
\end{equation}
is the density operator in position space. Eq.~\eqref{eq:continuityposition} immediately implies Eq.~\eqref{eq:interactioncontinuity} in momentum space, showing that the current is conserved for any finite-sized system with either open or periodic boundary conditions. Eq.~\eqref{eq:jIposition} is reminiscent of the Irving-Kirkwood form of the stress tensor in systems with instantaneous pair interactions~\cite{irving1950statistical,bradlyn2012kubo}. As in that case, here we must make a choice about which path current flows along between lattice sites; the function $\mathbf{s}_{ij,\alpha}$ encodes this choice. The fact that different choices of path lead to physically distinct conserved currents is reminiscent of N\"other's second theorem: the difference between current operators corresponding to different choices of path is purely transverse and divergence-less off-shell\cite{phillips2019nother}.

\subsection{Current operator in Fourier space}\label{sec:currentft}
Next, let us Fourier transform $\mathbf{j}_\mathrm{I}(\mathbf{r})$ to obtain 
\begin{equation}\label{eq:ftdef}
\mathbf{j}_{\mathrm{I},\mathbf{q}} = \int d\mathbf{r} e^{-i\mathbf{q}\cdot\mathbf{r}}\mathbf{j}_{\mathrm{I}}(\mathbf{r}).
\end{equation}
We find immediately that, for any finite-sized system with either open or periodic boundary conditions,
\begin{widetext}
\begin{align}
&j_{\mathrm{I},\mathbf{q}}^\mu =i\sum_{\substack{\alpha\beta\\\mathbf{R}_1\mathbf{R}_2\\\mathbf{R}_3\mathbf{R}_4}} U_{\alpha\beta}(\mathbf{R}_1+\mathbf{R}_3-\mathbf{R}_2-\mathbf{R}_4)c^\dag_{\alpha\mathbf{R}_1}c_{\alpha\mathbf{R}_2}c^\dag_{\beta\mathbf{R}_3}c_{\beta\mathbf{R}_4}\times\int_0^1d\lambda s^{\prime,\mu}_{12,\alpha}(\lambda)e^{-i\mathbf{q}\cdot\mathbf{s}_{12,\alpha}(\lambda)}+s^{\prime,\mu}_{34,\beta}(\lambda)e^{-i\mathbf{q}\cdot\mathbf{s}_{34,\beta}(\lambda)}\label{eq:jIq}
\end{align}

\end{widetext}

Several general remarks are in order before we proceed to analyze Eq.~\eqref{eq:jIq}. First, we note that we can also use Eq.~\eqref{eq:mincoupling} to take a second variational derivative of the Hamiltonian before setting $\mathbf{A}\rightarrow 0$. In this way we can derive the diamagnetic contribution to the current. While its explicit form is not particularly illuminating, we note that it is nonlocal due to the presence of two integrals over paths $\mathbf{s}_{ij,\alpha}$. This nonlocality leads precisely to the anomalous $f$-sum rule Eq.~\eqref{eq:OHK_DC}. 

Second, we note that the allowed values of the wavevector $\mathbf{q}$ depend on the choice of boundary conditions of the finite system. From Eq.~\eqref{eq:ftdef}, we see that for a system with periodic boundary conditions $\mathbf{r}\sim\mathbf{r}+\mathbf{L}$, the periodicity of $\mathbf{j}_{I}(\mathbf{r})$ requires that $\mathbf{q}$ be quantized via the standard relationship 
\begin{equation}\label{eq:pbcquantization}
\mathbf{q}=2\pi(n_x/L_x,n_y/L_y,n_z/L_z)    
\end{equation}
where $n_x,n_y,n_z\in \mathbb{Z}$ 
are integers. With open boundary conditions, on the other hand, there is no restriction on the allowed values of $\mathbf{q}$. This is because $\mathbf{q}$ is not the wavevector of any momentum-carrying excitation, but is simply the fourier component of $\mathbf{j}_{I}(\mathbf{r})$ (one can view it as the wavevector of an external electric field to which the system is coupled, which has no constraints with open boundary conditions).

Third, we again emphasize that our derivation of the current is valid for any finite-sized system with either periodic or open boundary conditions. Care must be taken to apply Eqs.~\eqref{eq:jIposition} or \eqref{eq:jIq} directly in the infinite-sized (thermodynamic) limit. 
The HK Hamiltonian \eqref{eq:ohkposition} contains a thermodynamically large number of terms, since it involves correlated hopping processes that couple sites at all positions. When we minimally couple the interaction to an external vector potential via Eq.~\eqref{eq:mincoupling}, this means that the vector potential at point $\mathbf{r}$ enters into a thermodynamically large number of terms in the Hamiltonian. This poses a problem if we try to define the current directly in the limit that the system size $L\rightarrow\infty$ limit: to derive Eq.~\eqref{eq:jIposition} we took the derivative of the Hamiltonian term-by-term, which is not guaranteed to converge to the derivative of the Hamiltonian as a whole if there are an infinite number of nonvanishing terms in the Hamiltonian. As such, our definition of the current operator in minimal coupling requires that we work in a finite-sized system of length $L$, taking the limit $L\rightarrow\infty$ at the end of any calculation. We will adopt the prescription that to consistently define the current, we should work in a finite-size system and take the limit as the system size goes to infinity after evaluating derivatives and commutators.  For a finite-sized system, there are a finite number of terms in the sum and we can compute the variation derivative in Eq.~\eqref{eq:jIposition} term-by-term as we have done.


This issue comes to the foreground when we consider the $\mathbf{q}=0$ limit of the current $\mathbf{j}_{\mathrm{I},\mathbf{q}}$. Here we consider open boundary conditions and periodic boundary conditions separately. For a finite system with open boundary conditions, $\mathbf{q}$ can take any value, so we can evaluate $\lim_{\mathbf{q}\rightarrow 0} j^\mu_{\mathrm{I},\mathbf{q}}$. For a finite-size system, we can evaluate the limit term-by-term since there are only a finite number of terms in the sum. To do so, we can Taylor expand Eq.~\eqref{eq:jIq} for small $\mathbf{q}$ and use our definition of $\mathbf{s}_{ij,\alpha}$ to find
\begin{widetext}
\begin{align}
j^\mu_{\mathrm{I},\mathbf{q}\rightarrow 0} \sim i\sum_{\substack{\alpha\beta\\\mathbf{R}_1\mathbf{R}_2\\\mathbf{R}_3\mathbf{R}_4}} &U_{\alpha\beta}(\mathbf{R}_1+\mathbf{R}_3-\mathbf{R}_2-\mathbf{R}_4)c^\dag_{\alpha\mathbf{R}_1}c_{\alpha\mathbf{R}_2}c^\dag_{\beta\mathbf{R}_3}c_{\beta\mathbf{R}_4}\times[\mathbf{R}_2+\mathbf{R}_4-\mathbf{R}_1-\mathbf{R}_3 + \mathcal{O}(\mathbf{q})].\label{eq:jIq0open}
\end{align}
\end{widetext}

For the case of a $\mathbf{k}$-independent OHK interaction, the leading order terms in the sum each vanish [see our discussion under Eq.~\eqref{eq:ohkposition}]. This shows that the current operator is regular as $\mathbf{q}\rightarrow 0$ for any finite-sized system with open boundary conditions and $\mathbf{k}$-independent OHK interaction. Note, however, that our derivation requires that $\mathbf{q}$ be sufficiently small that $\mathbf{q}\cdot\mathbf{s}_{ij,\alpha}(\lambda)$ is small for all choices of $i,j,\alpha$ and $\lambda$. This requires $\mathbf{q}$ to be smaller than the inverse linear dimension of the system. As such, the $\mathbf{q}\rightarrow 0$ limit and the thermodynamic limit do not commute with each other for systems with infinitely long-ranged interactions. We must take care then to be explicit about when the thermodynamic limit is taken in any computation of observables in the HK model. This subtlety was overlooked in a previous paper\cite{guerci2024electrical}.

The same conclusion holds true with periodic boundary conditions, with the caveat that with periodic boundary conditions and finite size, we cannot take the limit $\mathbf{q}\rightarrow 0$ since $\mathbf{q}$ takes only discrete values. Instead, we can evaluate $j^\mu_{I,\mathbf{q}=0}$ directly from Eq.~\eqref{eq:jIq} [or alternatively, by integrating Eq.~\eqref{eq:jIposition} over position]. Either way, we find with periodic boundary conditions
\begin{widetext}
\begin{align}
j^\mu_{I,\mathbf{q}= 0} = i\sum_{\substack{\alpha\beta\\\mathbf{R}_1\mathbf{R}_2\\\mathbf{R}_3\mathbf{R}_4}} &U_{\alpha\beta}(\mathbf{R}_1+\mathbf{R}_3-\mathbf{R}_2-\mathbf{R}_4)c^\dag_{\alpha\mathbf{R}_1}c_{\alpha\mathbf{R}_2}c^\dag_{\beta\mathbf{R}_3}c_{\beta\mathbf{R}_4}\times[\mathbf{f}(\mathbf{R}_1,\mathbf{R}_2)+\mathbf{f}(\mathbf{R}_3,\mathbf{R}_4 )],\label{eq:jIq0periodic}
\end{align}
\end{widetext}

where $\mathbf{f}(\mathbf{R}_i,\mathbf{R}_j)$ is the periodization of the displacement vector $\mathbf{R}_j-\mathbf{R}_i$. Since $\mathbf{f}(\mathbf{R}_i,\mathbf{R}_j)$ can be taken to vanish whenever $\mathbf{R}_j-\mathbf{R}_i$ is a multiple of the system size, Eq.~\eqref{eq:jIq0periodic} vanishes for the momentum-independent OHK interaction as required.

 As an aside, note additionally that for a uniform transverse electric field in periodic boundary conditions, the electromagnetic gauge field is spatially uniform (gauge equivalent to twisted boundary conditions). For a uniform gauge field $A_\mu = \alpha_\mu$, the Peierls phase factors appearing in Eq.~\eqref{eq:mincoupling} can be evaluated to yield
\begin{equation}
\begin{split}
&\int_{\mathbf{R}_1+\mathbf{r}_\alpha}^{\mathbf{R}_2+\mathbf{r}_\alpha}\mathbf{A}(\mathbf{x})\cdot d\mathbf{x} +\int_{\mathbf{R}_3+\mathbf{r}_\beta}^{\mathbf{R}_4+\mathbf{r}_\beta} \mathbf{A}(\mathbf{x})\cdot d\mathbf{x} \\
&= \alpha_\mu[f^\mu(\mathbf{R}_1,\mathbf{R}_2) +f^\mu(\mathbf{R}_3,\mathbf{R}_4) ].
\end{split}
\end{equation}
This means that a uniform vector potential does not enter into the interaction part of the Hamiltonian for the momentum-independent HK interaction, similar to our discussion following Eq.~\eqref{eq:jIq0periodic}. Thus, the interaction contribution to the current density is not probed by the sensitivity to twisted boundary conditions that lead to the uniform DC Hall conductivity in the framework of Niu, Thouless, and Wu\cite{ntw}. Hence, for the purpose of calculating the Hall conductivity in the $\mathbf{q}\rightarrow 0$ limit---and hence the Chern number of the ground state---one can neglect $\mathbf{j}_{\mathrm{I},\mathbf{q}}$\cite{sounak_tbd}.

\subsection{Current operator in the thermodynamic limit}\label{sec:thermo}

To further explore this subtlety of the thermodynamic limit, it is helpful to choose an explicit form for the functions $\mathbf{s}_{ij,\alpha}(\lambda)$. We take for simplicity the geodesic path
\begin{equation}
\mathbf{s}_{ij,\alpha}(\lambda) = \mathbf{R}_i+\mathbf{r}_\alpha + \lambda\mathbf{f}(\mathbf{R}_i,\mathbf{R}_j),
\end{equation}
where for open boundary conditions
\begin{equation}
    \mathbf{f}(\mathbf{R}_i,\mathbf{R}_j) = \mathbf{R}_j-\mathbf{R}_i,
\end{equation}
and for periodic boundary conditions we take $\mathbf{f}(\mathbf{R}_i,\mathbf{R}_j)$ to be the periodization of the shortest displacement vector between $\mathbf{R}_j$ and $\mathbf{R}_i$ (which has a discontinuity when $|R_i^\mu - R_j^\mu| >L^\mu/2$, with $L^\mu$ the linear extent of the system in the $\mu$ direction) as introduced above. With this choice, we can carry out the integration over $\lambda$ to find
\begin{widetext}
\begin{align}
   & j^\mu_{I,\mathbf{q}\neq 0}=-\sum_{\substack{\alpha\beta\\\mathbf{R}_1\mathbf{R}_2\\\mathbf{R}_3\mathbf{R}_4}} U_{\alpha\beta}(\mathbf{R}_1+\mathbf{R}_3-\mathbf{R}_2-\mathbf{R}_4)c^\dag_{\alpha\mathbf{R}_1}c_{\alpha\mathbf{R}_2}c^\dag_{\beta\mathbf{R}_3}c_{\beta\mathbf{R}_4}\times\nonumber\\
   &\times\left[ \frac{f^\mu(\mathbf{R}_1,\mathbf{R}_2)}{\mathbf{q}\cdot \mathbf{f}(\mathbf{R}_1,\mathbf{R}_2)}e^{-i\mathbf{q}\cdot\mathbf{r}_\alpha}(e^{-i\mathbf{q}\cdot \mathbf{R}_2}-e^{-i\mathbf{q}\cdot \mathbf{R}_1}) +\frac{f^\mu(\mathbf{R}_3,\mathbf{R}_4)}{\mathbf{q}\cdot \mathbf{f}(\mathbf{R}_3,\mathbf{R}_4)}e^{-i\mathbf{q}\cdot\mathbf{r}_\beta}(e^{-i\mathbf{q}\cdot \mathbf{R}_4}-e^{-i\mathbf{q}\cdot \mathbf{R}_3})\right]\label{eq:jIqspecific}.
\end{align}
\end{widetext}

Eq.~\eqref{eq:jIqspecific} agrees with the interaction contribution to the current derived directly from the continuity equation \eqref{eq:interactioncontinuity}, which is therefore satisfied for all $\mathbf{q}$. We thus verify in momentum space that the minimal coupling current is conserved, and that the continuity equation \eqref{eq:interactioncontinuity} is valid even for the HK interaction contrary to previous claims\cite{guerci2024electrical}.

At first glance, it may appear that Eq.~\eqref{eq:jIqspecific} is singular as $\mathbf{q}\rightarrow 0$ and therefore inconsistent with Eqs.~\eqref{eq:jIq0open} and \eqref{eq:jIq0periodic} for the $\mathbf{q}=0$ Fourier component of the current. However, the apparent singularity at $\mathbf{q}=0$ is a removable discontinuity. This is especially clear in the case of open boundary conditions, where we can take $\mathbf{q}\rightarrow 0$ continuously. for $|\mathbf{q}|\ll 1/L$ with $L$ the linear size of the system, we can make use of
\begin{equation}
    \lim_{x\rightarrow 0 }\frac{e^{-ix}-1}{x} = -i 
\end{equation}
to immediately recover Eq.~\eqref{eq:jIq0open},
\begin{equation}
    \lim_{\mathbf{q}\rightarrow 0} j^\mu_{I,\mathbf{q}\neq 0} = j^\mu_{I,\mathbf{q}\rightarrow 0}.\label{eq:limit}
\end{equation}
With periodic boundary conditions, we cannot take the limit $\mathbf{q}\rightarrow 0$ due to the quantization of $\mathbf{q}$ in Eq.~\eqref{eq:pbcquantization}. Attempting to naively set $\mathbf{q}=0$ via $n_x=n_y=n_z=0$ yields an indeterminant form (not a divergence) thereby signaling a removable discontinuity. To remove the discontinuity, we can assign a consistent definition to $j_{\mathrm{I},\mathbf{q}=0}$ using Eq.~\eqref{eq:jIq0periodic}. We thus see that with either open or periodic boundary conditions, the apparent singularity of the current at $\mathbf{q}=0$ is a removable discontinuity. 

\section{Relation to the Single impurity Anderson Model}
To see the relationship of these issues with DMFT, we  consider the physical interpretation of DMFT as a single-impurity Anderson model (SIAM) given by the Hamiltonian
\beq
H&=&\sum_{\mathbf k \sigma} n_{\mathbf k \sigma} \epsilon_{\mathbf k}+Un_{d\uparrow}n_{d\downarrow}+\sum_{\mathbf k \sigma}V_{\mathbf k} c^\dagger_{\mathbf k \sigma}a_{ d \sigma}+{\rm h.c.}\nonumber\\
&=&H_0+H_{\rm int}
\eeq
with a coupling $V_{\mathbf k}$ between an impurity state and a conduction electron created with operators $a_{ d \sigma}^\dagger$ and $c_{\mathbf k \sigma}^\dagger$, respectively.  As the interaction term in $H_0$ does not couple to the electromagnetic gauge field, it cannot contribute to the current.  Treating the conduction electron charge density $\rho_{\mathbf q}$ as before, we find that the commutator
\beq
[H_0,\rho_{\mathbf q}]=\sum_{\mathbf k \sigma} (\epsilon_{\mathbf {k +q}}-\epsilon_{\mathbf k}) c^\dagger_{\mathbf {k +q}\sigma}c_{\mathbf k \sigma}
\eeq
yields the standard term that vanishes in the limit of $q\rightarrow 0$. 
 Because the impurity and conduction electron operators act in different Hilbert spaces, they transform independently under $U(1)$.  As a consequence, the commutator

 


\beq
 [H_{\rm int},\rho_{\mathbf q}]=\sum_{\mathbf k} V_{\mathbf k} a_{d \sigma}^\dagger c_{ \mathbf {k +q} \sigma} - V_{\mathbf {k +q}} c_{\mathbf {k }\sigma}^\dagger a_{ d \sigma}
 \eeq

 does not vanish in the limit $q\rightarrow 0$ which will necessarily result in a non-vanishing of the double commutator in Eq. (\ref{eq:double_commutator}).   As a consequence, the single-impurity reduction of DMFT has the same pitfall as does HK.  However, there is a subtlety here which if we fix can make the connection even tighter.  Namely, we have ignored the charge density at the impurity. Assuming that every conduction electron can hop to the impurity (i.e. that we are approaching the infinite-dimensional limit in which the DMFT mapping is exact), then the correct total density is of the form
 \beq
 \label{rhoq}
 \rho_\mathbf{q}\rightarrow \sum_k c^\dagger_{\mathbf k,\sigma} c_{\mathbf {k+q},\sigma}+\delta_{\mathbf{q},0} a^\dagger_{d\sigma} a_{d\sigma}.
 \eeq
 Clearly, if we set $\mathbf{q}=0$, we obtain two contributions.  However, if we restrict to $\mathbf{q}\ne 0$ and then take the limit, only the first term survives.  This is identical to the problem with the HK model.  In essence, the Anderson impurity model in this limit has contributions from an infinite range as any electron regardless of its location can interact with the impurity.  Consequently, for the Anderson impurity problem, $\lim_{\mathbf{q}\rightarrow 0}\rho_\mathbf{q}\ne\rho_{\mathbf{q}=0}$.  As in HK, the discontinuity at $\mathbf{q}=0$ is still removable and the procedure outlined for defining the gauge-invariant current works here as well. We can also relax the form of Eq. (\ref{rhoq}) by replacing $\delta_{\mathbf{q},0}$ with any $f(\mathbf{q})$ such that $f(\mathbf{q}=0)=1$, which corresponds to limiting the range of the hybridization between the impurity and the conduction electrons.  From the point of view of HK, this would correspond to limiting the range of the HK interaction in position space thereby removing the long-range contribution to the diamagnetic current and $f$-sum rule.  For cluster methodology, increasing the size of the cluster is analogous.  Strictly for the ultra-local limit of a single impurity in which no momentum dependence appears in the self-energy, $n=1$ HK and DMFT have an intimate connection. In the $d\rightarrow\infty$ limit of the SIAM, both the matrix element and the hybridization  between the impurity and the conduction band are scaled by $1/\sqrt{d}$.  In the SIAM, band states with momentum $k$ and $k'$ are mixed by $V_{kd}V^*_{dk'}/\epsilon(k)$.  Because of the $d\rightarrow\infty$ scaling $V_{kd}\rightarrow V_{kd}/\sqrt{d}$ and $t\rightarrow t/\sqrt{d}$, the transition rate due to impurity scattering between states with momentum $\mathbf{k}$ and $\mathbf{k}'$ vanishes as $d\rightarrow\infty$.  Consequently in $d=\infty$, no mixing survives and all the band momentum states are decoupled. Note, however, that since the density of states diverges as $\sqrt{d}$ in infinite dimensions, the electronic self-energy correction due to impurity scattering remains finite and momentum independent. Hence, just as in HK, DMFT in infinite dimensions is a theory of interacting electrons in decoupled momentum sectors. 

This is not a surprise, as it has recently been shown\cite{mai2024} that the convergence of the $n^d$-orbital extension of HK to the Hubbard model scales as $1/n^{2\gamma(d)}$, where $\gamma(d)$ is unity in $d=1$ increases from there.  In $d=\infty$, all the fluctuations vanish for $n>1$, implying the universality class for the Mott transition in both cases is identical.  Aside from the scaling argument on the fluctuations and the decoupling of the scattering between the momenta,  both HK and DMFT, though for different reasons, have a central peak (band overlap for HK but single-impurity physics for DMFT) in the density of states as the Mott insulating state is approached.   All state-of-the-art simulations\cite{tmaier} find that in any finite dimension, the central peak does not survive but still the Mott transition exists.  Consequently, the quasiparticle peak of DMFT is ancillary to Mottness as is the band overlap in HK.  

\section{Concluding Remarks}

We have reported the density-density susceptibility as well as the conserved current operator for HK and OHK models.  The key features that arise from the two-pole structure of the density-density response are 1) mixing between the upper and lower Hubbard bands in the particle-hole continuum, 2) a plasma frequency that diverges inversely with the momentum, and 3) a lack of commutativity of the long-wavelength and thermodynamic limits.  The mixing between the upper and lower Hubbard bands is tied to the zeros in the Green function.  As long as the zeros are present, the HK represents a fixed point which is stable to all local repulsive interactions.  Consequently, even if Hubbard were included, no qualitatively new physics would arise.  This has been verified recently\cite{kunyang} by the inclusion of local Hubbard interactions which were shown only to yield an additive correction to the imaginary part of the self energy.  Since the self-energy already diverges as a result of the zero-structure of the Green function, the additive correction does not provide any qualitatively new physics.  

The lack of commutativity stems from the subtlety in taking the $\mathbf{q}=0$ limit of the double commutator of the density with the Hamiltonian which was ignored in previous work\cite{guerci2024electrical}.  We have derived the charge current from minimal coupling for open and periodic boundary conditions both of which are gauge invariant and identical in form.  We show that regardless of boundary conditions the current satisfies the continuity equation in momentum space when the thermodynamic limit is carefully taken, avoiding criticisms noted previously\cite{guerci2024electrical}. The anomalous double commutator instead signifies that the diamagnetic current in HK models is nonlocal in position space. Consequently, transport properties can be formulated cleanly in OHK models. These issues will be explored in more detail in a forthcoming work.

\textbf{Acknowledgements} 
B.B. thanks Y. Guan and S. Sinha for discussions on a related project that led to Eq.~\eqref{eq:mincoupling}. This work was supported by the Center for Quantum Sensing and Quantum Materials, a DOE Energy Frontier Research Center, grant DE-SC0021238 (P.M., B.B., and P.W.P.). B.B. received additional support from the Alfred P. Sloan Foundation, and the National Science Foundation under grant DMR-1945058 for work on the OHK model generally.  PWP also acknowledges NSF DMR-2111379 for partial funding of the HK work which led to these results.  

\appendix
\section{Proof of factorization of density-density correlation function}

In this section, we prove Eq.~\eqref{chig} of the main text. That is, we show that the density-density correlation function $\chi^0_\mathrm{HK}(\mathbf{q}\neq 0,\tau)$ in the orbital HK model factorizes into a product of single-particle Green functions. We start by observing that the orbital HK Hamiltonian Eq.~\eqref{ohk} and the density operator $\rho_\mathbf{q}$ can be written respectively as
\begin{align}
H_{\mathrm{OHK}} &= \sum_\mathbf{k} H_\mathbf{k},\label{eq:hkfactorizes} \\
\rho_\mathbf{q} &= \sum_{\alpha\mathbf{k}}c^\dag_{\alpha\mathbf{k}}c_{\alpha\mathbf{k+q}}.
\end{align}
Next, we consider the definition of $\chi^0_\mathrm{HK}(\mathbf{q}\neq 0,\tau)$. Taking $\tau>0$ without loss of generality (this is all that is needed to compute the Fourier transform, and $\tau<0$ values can be reconstructed using periodicity), we have
\begin{align}
\chi^0_\mathrm{HK}(\mathbf{q}\neq 0,\tau>0) &= -\frac{1}{N}\langle T_\tau[\rho_\mathbf{q}(\tau)\rho_-\mathbf{q}(0)]\rangle\nonumber \\
&= -\frac{1}{N}\langle \rho_\mathbf{q}(\tau)\rho_-\mathbf{q}(0)\rangle\nonumber \\ 
& = -\frac{1}{N}\sum_{\alpha\beta\mathbf{k}\mathbf{k'}}\langle c^\dag_{\alpha\mathbf{k}}(\tau)c_{\alpha\mathbf{k+q}}(\tau)c^\dag_{\beta\mathbf{k'}}(0)c_{\beta\mathbf{k'-q}}(0)\rangle.\label{eq:chi_intermediate}
\end{align}
So far we have not made use of any properties of the HK Hamiltonian. We now observe the following facts:
\begin{enumerate}
    \item From Eq.~\eqref{eq:hkfactorizes}, it follows that time-evolved creation and annihilation operators $c_{\alpha\mathbf{k}}(\tau), c^\dag_{\alpha\mathbf{k}}(\tau)$ are linear combinations of operators at fixed $\mathbf{k}$.
    \item The HK density matrix $e^{-\beta H_\mathrm{OHK}}/Z$ factorizes as $\prod_\mathbf{k} e^{-\beta H_\mathbf{k}}/Z_\mathbf{k}$.
    \item The many-body eigenstates of $H_\mathbf{OHK}$ factorize as tensor products $\ket{N}=\otimes_\mathbf{k}\ket{n_\mathbf{k}}$ with energies $E_N = \sum_\mathbf{k} E_{n_\mathbf{k}}$.
\end{enumerate}

\begin{widetext}
Using property 1 and the anticommutation relations, we can reorder the creation and annihilation operators in Eq.~\eqref{eq:chi_intermediate} for $\mathbf{q}$ to find
\begin{equation}
\chi^0_\mathrm{HK}(\mathbf{q}\neq 0,\tau>0)=-\frac{1}{N}\sum_{\alpha\beta\mathbf{k}\mathbf{k'}}\langle c^\dag_{\alpha\mathbf{k}}(\tau)c_{\beta\mathbf{k'-q}}(0)c_{\alpha\mathbf{k+q}}(\tau)c^\dag_{\beta\mathbf{k'}}(0)\rangle.
\end{equation}
Next, we insert a complete set of states to get
\begin{align}
\chi^0_\mathrm{HK}&(\mathbf{q}\neq 0,\tau>0)=-\frac{1}{N}\sum_{\substack{\alpha\beta\mathbf{k}\mathbf{k'}\\A,B}}
\bra{A}\frac{e^{-\beta H_\mathrm{OHK}}} {Z}c^\dag_{\alpha\mathbf{k}}(\tau)c_{\beta\mathbf{k'-q}}(0)\ket{B}\bra{B}c_{\alpha\mathbf{k+q}}(\tau)c^\dag_{\beta\mathbf{k'}}(0)\ket{A}\label{eq:chi_with_intermediate_states}
\end{align}
Now note that from property 3 that we can write $\ket{A}=\otimes_{k}\ket{a_\mathbf{k}}$ and $\ket{B}=\otimes_\mathbf{k}\ket{b_\mathbf{k}}$. Since $H_\mathbf{OHK}$ preserves particle number at each $\mathbf{k}$, each of the states $\ket{a_\mathbf{k}}$ and $\ket{b_\mathbf{k}}$ has a fixed particle nubmer. This implies that the matrix elements in Eq.~\eqref{eq:chi_with_intermediate_states} vanish unless $\mathbf{k}=\mathbf{k'-q}$. This allows us to perform the sum over $\mathbf{k'}$. Furthermore, we can sum over all $\mathbf{k}$ sectors other than $\mathbf{k}$ and $\mathbf{k+q}$. Introducing the two-$\mathbf{k}$-sector states $\ket{a_\mathbf{k}a_\mathbf{k+q}} and \ket{b_\mathbf{k}b_\mathbf{k+q}}$ with energies $E_{a_\mathbf{k}}+E_{a_\mathbf{k+q}}$ and $E_{b_\mathbf{k}}+E_{b_\mathbf{k+q}}$ respectively, we have
\begin{align}
\chi^0_\mathrm{HK}(\mathbf{q}\neq 0,\tau>0)=-\frac{1}{N}\sum_\mathbf{k\alpha\beta}\sum_{a_{\mathbf{k}}a_{\mathbf{k+q}}b_{\mathbf{k}}b_{\mathbf{k+q}}}\frac{e^{-\beta E_{a_\mathbf{k}}}} {Z_\mathbf{k}}\frac{e^{-\beta E_{a_\mathbf{k+q}}}}{Z_\mathbf{k+q}}\bra{a_\mathbf{k}a_\mathbf{k+q}}c^\dag_{\alpha\mathbf{k}}(\tau)c_{\beta\mathbf{k}}(0)\ket{b_\mathbf{k}b_\mathbf{k+q}}\bra{b_\mathbf{k}b_\mathbf{k+q}}c_{\alpha\mathbf{k+q}}(\tau)c^\dag_{\beta\mathbf{k+q}}(0)\ket{a_\mathbf{k}a_\mathbf{k+q}}
\end{align}
Now since the first matrix element does not involve $\mathbf{k+\mathbf{q}}$, it enforces $a_\mathbf{k+q}=b_{\mathbf{k+q}}$. Similarly, the second matrix element enforces $a_\mathbf{k}=b_\mathbf{k}$. This allows us to perform the sums over $b_\mathbf{k}$ and $b_\mathbf{k+q}$ to find
\begin{align}
\chi^0_\mathrm{HK}(\mathbf{q}\neq 0,\tau>0)&=-\frac{1}{N}\sum_\mathbf{k\alpha\beta}\sum_{a_{\mathbf{k}}a_{\mathbf{k+q}}}\frac{e^{-\beta E_{a_\mathbf{k}}}} {Z_\mathbf{k}}\bra{a_\mathbf{k}a_\mathbf{k+q}}c^\dag_{\alpha\mathbf{k}}(\tau)c_{\beta\mathbf{k}}(0)\ket{a_\mathbf{k}a_\mathbf{k+q}}\frac{e^{-\beta E_{a_\mathbf{k+q}}}}{Z_\mathbf{k+q}}\bra{a_\mathbf{k}a_\mathbf{k+q}}c_{\alpha\mathbf{k+q}}(\tau)c^\dag_{\beta\mathbf{k+q}}(0)\ket{a_\mathbf{k}a_\mathbf{k+q}} \nonumber \\ 
&=-\frac{1}{N}\sum_\mathbf{k\alpha\beta}\sum_{a_{\mathbf{k}}a_{\mathbf{k+q}}}\frac{e^{-\beta E_{a_\mathbf{k}}}} {Z_\mathbf{k}}\bra{a_\mathbf{k}}c^\dag_{\alpha\mathbf{k}}(\tau)c_{\beta\mathbf{k}}(0)\ket{a_\mathbf{k}}\frac{e^{-\beta E_{a_\mathbf{k+q}}}}{Z_\mathbf{k+q}}\bra{a_\mathbf{k+q}}c_{\alpha\mathbf{k+q}}(\tau)c^\dag_{\beta\mathbf{k+q}}(0)\ket{a_\mathbf{k+q}}\nonumber \\ 
&= -\frac{1}{N}\sum_\mathbf{k\alpha\beta}\langle c^\dag_{\alpha\mathbf{k}}(\tau)c_{\beta\mathbf{k}}(0)\rangle\langle c_{\alpha\mathbf{k+q}}(\tau)c^\dag_{\beta\mathbf{k+q}}(0)\rangle \nonumber \\ 
&=-\frac{1}{N}\sum_\mathbf{k\alpha\beta}\langle T_\tau c^\dag_{\alpha\mathbf{k}}(\tau)c_{\beta\mathbf{k}}(0)\rangle\langle T_\tau c_{\alpha\mathbf{k+q}}(\tau)c^\dag_{\beta\mathbf{k+q}}(0)\rangle \\
&= \frac{1}{N}\sum_\mathbf{k}\mathrm{tr} [G_{\mathbf{k}}(-\tau) G_\mathbf{k+q}(\tau)]
\end{align}
where $G_\mathbf{k}(\tau)$ is the imaginary time-ordered two point function viewed as a matrix in the $\alpha,\beta$ orbital space, and we have used antiperiodicity in imaginary time. This is the desired result.

Our result shows that for $\mathbf{q}\neq 0$, the HK density-density response function $\chi^0_\mathrm{HK}(\mathbf{q}\neq 0,\tau)$ is given exactly in terms of a product of single-particle Green functions. Our derivation treats the interaction exactly, so there is no further renormalization of the correlation function or vertices. Our results are consistent with the Ward identity for the HK Hamiltonian, which can be expressed as an operator identity in terms of the current operator derived in Sec.~\ref{sec:current}; The continuity equation \eqref{eq:continuityposition} is equivalent to the Ward identity for the correlation functions.

Note that our derivation in this Appendix relied crucially on the assumption that $\mathbf{q}\neq 0$, which allowed us to reorder the operators in going from Eq.~\eqref{eq:chi_intermediate} to Eq.~\eqref{eq:chi_with_intermediate_states}. For $\mathbf{q}=0$, reordering the operators produces additional terms and the correlation function does not factorize. In this sense, the factorization of correlation functions in HK models is distinct from the usual "Wick's theroem" factorization in free fermion models.

\end{widetext}
\bibliography{suscep}

\end{document}